\documentclass[fleqn,usenatbib]{mnras}

\usepackage{newtxtext,newtxmath}

\usepackage[T1]{fontenc}

\DeclareRobustCommand{\VAN}[3]{#2}
\let\VANthebibliography\thebibliography
\def\thebibliography{\DeclareRobustCommand{\VAN}[3]{##3}\VANthebibliography}

\usepackage{graphicx}	
\usepackage{amsmath}	

\title[X-ray Study of Nuclear Structure in NLRGs]{Broadband X-ray Spectral Study of Nuclear Structure in Local Obscured Radio Galaxies}
      
\author[Nakatani et al.]{
Yuya Nakatani,$^{1}$\thanks{E-mail: nakatani@kusastro.kyoto-u.ac.jp}
Yoshihiro Ueda,$^{1}$
Claudio Ricci,$^{2,3}$
Koki Inaba,$^{1}$
Shoji Ogawa,$^{1,4}$
\newauthor
Kenta Setoguchi,$^{1}$
Ryosuke Uematsu,$^{1}$
Satoshi Yamada,$^{5}$
and Tomohiro Yoshitake$^{1}$
\\
$^{1}$Department of Astronomy, Kyoto University, Kitashirakawa-Oiwake-cho, Sakyo-ku, Kyoto 606-8502, Japan\\
$^{2}$Instituto de Estudios Astrof\'isicos, Facultad de Ingenier\'ia y Ciencias, Universidad Diego Portales, Av. Ej\'ercito Libertador 441, Santiago, Chile\\
$^{3}$Kavli Institute for Astronomy and Astrophysics, Peking University, Beijing 100871, People's Republic of China\\
$^{4}$Institute of Space and Astronautical Science (ISAS), Japan Aerospace Exploration Agency (JAXA) 3-1-1 Yoshinodai, Chuo-ku, Sagamihara, Kanagawa, Japan\\
$^{5}$RIKEN Cluster for Pioneering Research, 2-1 Hirosawa, Wako, Saitama 351-0198, Japan
}

\date{Accepted XXX. Received YYY; in original form ZZZ}

\pubyear{2023}

\begin{document}
\label{firstpage}
\pagerange{\pageref{firstpage}--\pageref{lastpage}}
\maketitle

\begin{abstract}

Radio galaxies are a key population to
understand the importance of relativistic jets in AGN feedback.
We present the results of a systematic, broadband X-ray spectral analysis of hard X-ray selected radio galaxies to investigate their nuclear structures.
In this study, we focus on the
seven most radio-loud,
X-ray obscured narrow line radio galaxies 
in the \textit{Swift}/BAT 70 month AGN catalog.
The spectra from 0.5 keV up to 66 keV obtained with \textit{Suzaku} and
\textit{NuSTAR} of six objects are newly analyzed here by utilizing
the X-ray clumpy torus model (XCLUMPY),
whereas we refer to Ogawa et al. (2021) 
for the results of Centaurus A.
We find that these radio galaxies have similar torus covering fractions
compared with radio quiet AGNs at the same Eddington ratios 
($-3 < \log \lambda_{\rm Edd} < -1$).
This result implies that (1) the torus structure is not an important factor that determines the presence of jets and (2) AGN jets have physically little effect on the torus.
\end{abstract}




\section{Introduction} \label{sec:intro}

The mass of a supermassive black hole (SMBH) at the center of galaxy
and that of its host-galaxy bulge
show a tight correlation in the local universe
($z\sim0$; see \citealt{Kormendy13} for a review). Such a tight correlation suggests that the
central SMBHs and host galaxies have ``coevolved'' by controlling
their respective growths.
During the active galactic nucleus (AGN) phase a SMBH gains mass
by accretion, which is the dominant channel of the overall SMBH
growth over cosmic time (e.g., \citealt{Marconi04}; \citealt{Ueda14}).
To reveal the physical origin of the coevolution, we need to
understand how matter is transferred onto SMBHs from galaxy scale (AGN
feeding) and how AGNs
affect the surrounding environments (AGN feedback; \citealt{Fabian12}).
The nuclear structure of AGNs, including the ``tori''
(see e.g., \citealt{RamosAlmeida&Rocci17} for a review), accretion disks,
and outflows (jets and winds),
can provide us with important clues to solve these fundamental questions.

X-ray observations are a powerful tool to probe into the AGN structure
because X-rays have strong penetrating power through dust and gas
concealing the central engine, and are sensitive to nearly all
material including dust and gas with wide ranges of temperature and
ionized stage. 
In particular, unique information can be obtained through X-ray reflection signatures produced in the circumnuclear material irradiated by the central source
(e.g., \citealt{George91}).
This reprocessed radiation is characterized by a hard continuum with a
hump at $\sim$30~keV and fluorescence lines including 
the prominent iron K$\alpha$ line.

Radio galaxies (RGs) are considered to be a key population to
understand the importance of relativistic jets in AGN feedback since they could have
a significant impact on star formation
(e.g., \citealt{Wagner12}). 
One basic question still remains unanswered: are there any distinct differences 
in the nuclear structure of RGs with respect to radio-quiet AGN, besides the presence or absence of jets?
Previous
X-ray studies of broad line radio galaxies (BLRGs) suggest that the accretion disks
seem to be truncated before reaching to the innermost stable circular
orbits (ISCOs; \citealt{Kataoka07,Larsson08,Sambruna09,Tazaki10,Sambruna11,Lohfink17}).
This may be different from those of radio-quiet
AGNs, where the accretion disks are considered to
extend down to the ISCOs in order to produce
the relativistically broadened reflection features (e.g.,
\citealt{Tanaka95}). Other models were proposed to
interpret these reflection features (e.g., \citealt{Miyakawa12}), however,
making the comparison non-trivial.
\cite{Tombesi14} performed a systematic search for ultra fast outflows
(UFOs) in RGs by utilizing \textit{XMM-Newton} and \textit{Suzaku} data, and estimated
that $50\pm20$\% of RGs show UFOs. Since this fraction is similar
to that found in radio-quiet AGNs, the authors concluded that the
presence of jets did not preclude that of winds in AGNs.

In this work, we focus on the structures of the tori in RGs and compare them with those of radio-quiet AGNs.
The torus is a key element to understand both feeding and feedback mechanisms in an AGN,
since it plays the role of a mass reservoir, spatially connecting the SMBH and its host
galaxy. At the same time, the torus could be shaped by feedback from the
central engine via disk winds (\citealt{Elitzur&Shlosman06,Fukumura10}) and
via radiation pressure from the central disk (\citealt{Wada15,Ricci17Nature}). 
It is still unclear whether the presence of powerful jets 
is controlled by, or affects, the torus structure.
Previous studies of broadband X-ray spectra of RGs
covering energies above 10~keV, focused on constraining their torus structure, are still very limited in sample size (e.g.,
\citealt{Tazaki11,Tazaki13,Lohfink15,LaMassa23}).
In these works, conventional reflection models, such
as \textsc{pexrav} (\citealt{Magdziarz&Zdziarski95}), or so-called
``smooth'' torus models assuming uniform densities within given
boundaries (e.g., \citealt{Ikeda09,Murphy&Yaqoob09,Balokovic18}) were employed to
model the reflection components.
In reality, however, matter distribution in the torus is known to be complex,
composed of a clumpy medium.
To tackle this issue, \citet{Tanimoto19} have developed the XCLUMPY
model, an X-ray spectral model for a clumpy torus
irradiated by an X-ray source, utilizing
the Monte Carlo simulation for Astrophysics and Cosmology (MONACO:
\citealt{Odaka11,Odaka16}) framework.
In XCLUMPY, the clumps are distributed according to power-law and normal profiles in the radial and angular directions, respectively. 
This model has been successfully applied to the X-ray spectra of a large number of
AGNs to constrain the torus geometry
(e.g., \citealt{Miyaji19,Tanimoto20,Tanimoto22,Ogawa21,Uematsu21,Yamada21}; Inaba et al, \emph{in prep}).

This paper reports the results of a systematic broadband (0.5--65 keV)
X-ray spectral analysis of seven
narrow line radio galaxies (NLRGs) in
the local universe, utilizing the currently best available data set
obtained with \textit{Suzaku} and \textit{NuSTAR}.
Our sample is taken from the list of the most radio-loud objects
in the \textit{Swift}/BAT 70-month AGN catalog that satisfy the criteria
summarized in Section~\ref{sec:sample}.
We use the XCLUMPY model to determine the
torus geometry of each AGN.
Here we newly analyze the spectra of six objects, whereas we refer to
\cite{Ogawa21} for the results of Centaurus A.
This is the first work that
uniformly applies a clumpy torus model to the X-ray
spectra of a statistically well-defined sample of NLRGs.
The paper is organized as follows: 
Section~\ref{sec:sample} describes the sample selection. 
Sections~\ref{sec:obs_data} and \ref{sec:x-ray}
explain the data reduction and X-ray spectral analysis for the six RGs.
Their torus properties are discussed in Section~\ref{sec:discussion}.
Section~\ref{sec:conclusion} summarizes the conclusions.
In Appendix~A, we compare the results of our spectral fitting with those of
previous works for the individual objects.
Throughout the paper, we adopt the solar abundances of \citet{Anders89} and assume
the following cosmological parameters: $H_0 = 70\ \rm{km}\ \rm{s}^{-1}\ \rm{Mpc}^{-1}$, $\Omega_{\rm{M}} = 0.3$, and $\Omega_{\Lambda}=0.7$. 
Errors on the spectral parameters correspond to 90\% confidence limits for a single parameter of interest, unless otherwise stated.

\section{SAMPLE SELECTION} \label{sec:sample}

Our sample finally consists of
seven NLRGs selected from the \textit{Swift}/BAT
70-month AGN catalog\footnote{http://swift.gsfc.nasa.gov/results/bs70mon/} (\citealp{Baumgartner13}) by the following criteria.
\footnote{
  We exclude Cygnus A from our sample, 
  to avoid uncertainties in the spectral modeling due to the large contamination 
  from the intracluster medium (\citealt{Reynolds15}), although it satisfies the selection criteria.
}
To select the most radio-loud AGNs in the parent \textit{Swift}/BAT sample, 
we first excluded all blazars, and then set a lower-limit to the radio loudness parameter, $\log \rm R > -2.8$,
where $\rm R$ is defined as the ratio between the 1.4 GHz flux 
$\nu f_{\nu (1.4 \rm GHz)}$ in units of mJy Hz ($f_\nu$ is the flux density at the frequency $\nu$)
and the intrinsic 14--150~keV flux $f_{14-150 \rm keV}$
in units of J m$^{-2}$ s$^{-1}$
[$\rm R \equiv \log (\nu f_{\nu(1.4 \rm GHz)}/f_{14-150 \rm keV})$].
The radio-loudness was estimated for the AGN from the \textit{Swift}/BAT 70-month catalogue by considering literature values for their
total 1.4~GHz emission from the core and jets/lobes. 
For most of the sources these values were taken from the NRAO VLA Sky Survey (NVSS; \citealt{Condon98}) or from the Green Bank 1.4~GHz Northern Sky Survey (\citealt{White&Becker92}). 
For the sources of our sample, the 1.4~GHz fluxes we used were reported 
by the Green Bank survey (VII~Zw~292, 3C~403, 3C~105, 3C~452; \citealt{White&Becker92}) 
, by NVSS (\citealt{Condon98} for PKS~0326--288 and \citealt{Kuzmicz18} for PKS~2356--61)
, and by a VLA observation (Centaurus~A; \citealt{Condon96}).
The intrinsic 14--150~keV flux were taken from \cite{Ricci17apj}.

We further limit our sample
to X-ray obscured AGNs with line-of-sight absorption $\log (N^{\rm
  LOS}_{\rm H}/\rm cm^{-2}) > 23$, where $N^{\rm LOS}_{\rm H}$ was
obtained by \cite{Ricci17apj}.  This is because (1) we can more
reliably determine the torus parameters in heavily obscured AGNs than in unobscured AGNs
by utilizing information of
the line-of-sight absorption (\citealt{Ogawa21}) and
(2) the contribution of emissions from the relativistic jets in the
observed X-ray spectra is much smaller in edge-on cases (obscured
AGNs) than in face-on cases (unobscured AGNs) assuming that such jets are
perpendicular to the torus equatorial plane (see Section~\ref{subsec:Implications}). 
The lower threshold for $N^{\rm LOS}_{\rm H}$ is also applied to ensure that the
absorption is caused by torus matter, not by interstellar medium in
the host galaxy. We finally require that the objects were observed with
\textit{NuSTAR} to utilize high-quality hard X-ray spectra above 10~keV. 
The basic information of our sample
is listed in Table~\ref{tab:obj}.

\begin{figure}
    \includegraphics[width=8.5cm]{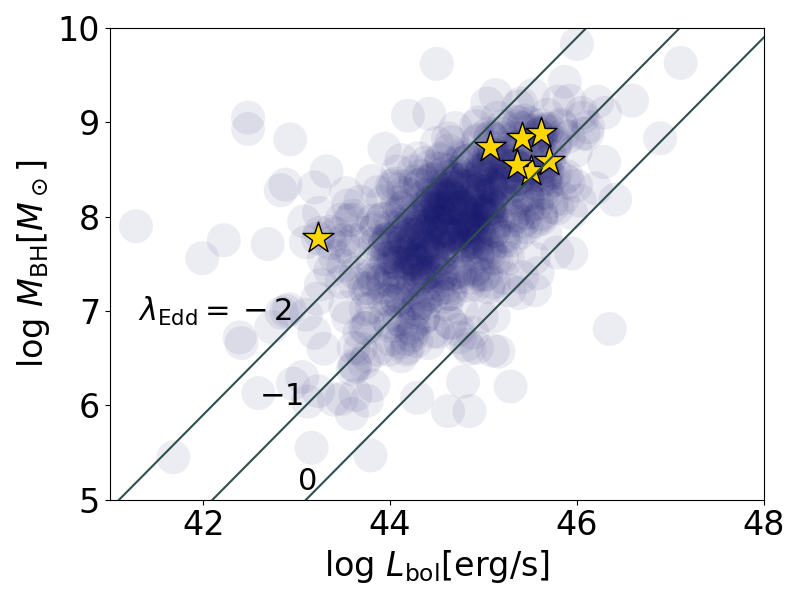}
    \caption{Relation of $M_\mathrm{BH}$ and $L_\mathrm{bol}$ reported in \citep{Koss22} for our sample (yellow stars; the left-most one is Centaurus~A) and for the whole \textit{Swift}/BAT 70-month sample (blue circle). Blazars are excluded. The black lines correspond to
constant Eddington-ratio lines: $\log \lambda_{\rm{Edd}} = -2, -1$, and 0.
}
\label{fig:Sample}
\end{figure}

Figure~\ref{fig:Sample} shows the relation between the black hole mass
$M_\mathrm{BH}$ and the bolometric luminosity $L_\mathrm{bol}$ for our
sample and the total \textit{Swift}/BAT 70-month AGN sample. As noticed,
the six objects other than Centaurus A (left-most object) are
distributed in a narrow region on this plane,
having relatively large black hole masses ($\log M_\mathrm{BH}[M_{\odot}]=8-9$)
and bolometric luminosities ($\log L_\mathrm{bol}[\mathrm{ergs/s}] = 45-46$) 
compared with the parent sample. They 
show Eddington ratios of $-2<\log \lambda_{\rm Edd}<-1$.
\begin{table*}
\caption{Summary of Objects. 
\label{tab:obj}
\newline
(1) Galaxy name. 
(2) Swift source name.
(3) Redshift from the NASA/IPAC Extragalactic Database (NED; https://ned.ipac.caltech.edu/).
(4) total Galactic $HI$ and $H_{2}$ values in units of 10$^{22}$ cm$^{-2}$ \protect\citep{Willingale13}.
(5) Radio loudness defined as the ratio between the 1.4 GHz flux and the intrinsic 14--150 keV flux.
(6) Radio AGN classification from the NED.
(7) 1.4~GHz flux density in units of mJy.
(8) logarithmic column density along the line-of-sight in units of cm$^{-2}$ \protect\citep{Ricci17apj}.
(9) References of column (3), (6), and (7). ``$\cdots$'' denotes no reference.
\protect\newline \protect\textbf{References.}
(B) \protect\cite{Bottinelli92}; (C96) \protect\cite{Condon96}; (C98) \protect\cite{Condon98}; (H) \protect\cite{Hardcastle99}; 
(J) \protect\cite{Jones09}; (Ki) \protect\cite{Kim21}; (Ku) \protect\cite{Kuzmicz18}; 
(L) \protect\cite{Loveday96}; (O) \protect\cite{Owen&Laing89}; (R) \protect\cite{Ricci17apj}; (S) \protect\cite{Spinrad85}; (W) \protect\cite{White&Becker92} 
}
\begin{tabular}{|l|l|l|l|c|l|l|l|l}
\hline
(1) & (2) & (3) & (4) & (5) & (6) & (7) &(8)& (9)\\
Object & Swift ID & Redshift &$N^{\mathrm{Gal}}_{\mathrm{H}}$ & Radio loudness& Radio Class & Radio flux
&$\log N^{\mathrm{LOS}}_{\mathrm{H}}$&Ref\\
\hline
PKS~0326--288	&SWIFT J0328.4-2846	&0.10877	&0.0102	&-2.8	&$\cdots$	&1450	&23.82	&(J,$\cdots$,C98)	\\
VII~Zw~292	&SWIFT J0950.5+7318	&0.0581	&0.0233	&-2.69	&FR II	&2653	&23.79	&(R,O,W)	\\
3C~403	&SWIFT J1952.4+0237	&0.0584	&0.195	&-2.69	&FR II	&5798	&23.69	&(Ki,H,W)	\\
3C~105	&SWIFT J0407.4+0339	&0.089	&0.22	&-2.63	&FR II	&5339	&23.75	&(S,H,W)	\\
Centaurus~A &SWIFT J1325.4-4301 & 0.00183 & 0.117 & -2.51 & $\cdots$& 278000& 23.02& (B,$\cdots$,C96)\\
3C~452	&SWIFT J2246.0+3941	&0.0811	&0.144	&-2.33	&FR II	&10197	&23.76	&(R,H,W)	\\
PKS~2356--61	&SWIFT J2359.3-6058	&0.09631	&0.0157	&-1.52	&$\cdots$	&26240	&23.16	&(L,$\cdots$,Ku)	\\
\hline
\end{tabular}
\end{table*}
\section{OBSERVATIONS AND DATA REDUCTION} \label{sec:obs_data}

Table~\ref{tab:obs} summarizes the \textit{Suzaku} and \textit{NuSTAR} data
of the six RGs (other than Centaurus A) analyzed here. 
The data reduction was conducted in accordance with the procedures described below.

\begin{table*}
\caption{Summary of X-ray Observations. \label{tab:obs}
\newline
Exposures are based on the good time intervals of XIS 0 for \textit{Suzaku} and those of FPMA for \textit{NuSTAR}.
}
\begin{tabular}{llcccc}
\hline
Object          & Observatory       &
ObsID           & Start Date (UT)   &
End Date (UT)   & Exposure (ks)
\\
\hline
PKS~0326--288	&\textit{Suzaku}	&704039010	&2010 Jan 30 21:20	&2010 Feb 01 07:50	&58	\\
~	&\textit{NuSTAR}	&60160155002	&2020 Jan 26 07:56	&2020 Jan 26 18:51	&22	\\
VII~Zw~292	&\textit{Suzaku}	&703018010	&2008 Nov 16 22:10	&2008 Nov 18 11:52	&81	\\
~	&\textit{NuSTAR}	&60160374002	&2016 Dec 05 02:51	&2016 Dec 05 09:06	&13	\\
3C~403	&\textit{Suzaku}	&704011010	&2009 Apr 08 19:53	&2009 Apr 10 02:35	&48	\\
~	&\textit{NuSTAR}	&60061293002	&2013 May 25 18:06	&2013 May 26 05:01	&20	\\
3C~105	&\textit{Suzaku}	&702074010	&2008 Feb 05 12:52	&2008 Feb 07 02:00	&38	\\
~	&\textit{NuSTAR}	&60061044002	&2013 Feb 15 02:06	&2013 Feb 15 04:51	&5	\\
~	&~	&60061044004	&2013 Feb 15 15:01	&2013 Feb 15 18:01	&6	\\
~	&~	&60061044006	&2013 Feb 16 03:56	&2013 Feb 16 06:36	&6	\\
3C~452	&\textit{Suzaku}	&702073010	&2007 Jun 16 09:30	&2007 Jun 17 19:54	&67	\\
~	&\textit{NuSTAR}	&60261004002	&2017 May 01 00:56	&2017 May 02 03:51	&51	\\
PKS~2356--61	&\textit{Suzaku}	&801016010	&2006 Dec 06 17:08	&2006 Dec 09 05:11	&101	\\
~	&\textit{NuSTAR}	&60061330002	&2014 Aug 10 17:16	&2014 Aug 11 05:06	&23	\\
\hline
\end{tabular}
\end{table*}

\subsection{\textit{Suzaku}} \label{subsec:Suzaku}

\textit{Suzaku} \citep{Mitsuda07} carried four X-ray CCD cameras
called the X-ray Imaging Spectrometers (XISs), which were sensitive to
X-rays in the 0.2--12~keV range. XIS0, XIS2, and XIS3 were frontside-illuminated
CCD cameras (XIS-FI), and XIS1 was backside-illuminated one (XIS-BI). 
\textit{Suzaku} also carried the hard X-ray detector (HXD),
a non-imaging instrument consisting of
p-i-n type silicon
photodiodes (hereafter PIN) and
gadolinium silicate [$\rm Gd_2 \rm Si \rm O_5(\rm Ce)$]
phoswich counters (hereafter GSO),
which were sensitive to hard X-rays with energies
of 10--70~keV and 40--600~keV, respectively. In this work, we only
utilize HXD-PIN data because our targets were too faint to be detected
with HXD-GSO.

The \textit{Suzaku}/XIS data were reduced by using the latest versions of the calibration database (CALDB) 
released on 2018 October 10 and HEASoft v6.30.1 packages.
We accumulated photon events in circular regions around the target with radii of
1--2 arcmin, depending on the source flux. 
The background spectra were extracted from circular regions 
with radii of 1.5--4 arcmin that were free of any obvious point sources.
The response matrix files (RMFs) and ancillary response files (ARFs)
were generated with the \textsc{XISRMFGEN} and \textsc{XISSIMARFGEN} tools \citep{Ishisaki07},
respectively. 
We combined the spectra and responses of XIS-FI using the \textsc{ADDASCASPEC} script.
The spectral bins were merged to contain at least 25 photons per each
bin in order to facilitate the use of $\chi^2$-statistics.

The HXD-PIN data were reduced by using the \textsc{AEPIPELINE} script. For
background subtraction we used the ``tuned'' non X-ray background
(NXB) event files \citep{Fukazawa09} to produce the NXB spectra, to
which simulated cosmic X-ray background (CXB) spectrum were added. In
the spectral analysis, we utilized only the energy range within
16--40~keV where the source flux is brighter than 3\% of the NXB level
(the maximum systematic error in the 15--70~keV range;
\citealt{Fukazawa09}).

\subsection{\textit{NuSTAR}} \label{subsec:NuSTAR}

\textit{NuSTAR} \citep{Harrison13} carries two focal plane modules
(FPMs: FPMA and FPMB). The FPMs, coupled with the multilayer coated
hard X-ray mirrors, are sensitive to X-ray energies of 3--79~keV. The
FPMs data were reprocessed by using HEAsoft v6.30.1 with CALDB version
20200429. The source spectra were extracted by considering circles with radii of 90--120 arcsec
(depending on the flux) centered on the target. The background spectra
were extracted from source-free circular regions with radii of 90--120
arcsec. The response matrix files (RMFs) and the ancillary response
files (ARFs) were generated with the \textsc{NUPRODUCTS} script.
We combined the spectra and responses of the two FPMs by using
\textsc{ADDASCASPEC} to achieve the highest signal-to-noise ratios in a single spectrum;
we confirmed that consistent results were obtained by 
fitting the FPMA and FPMB spectra separately.
For the case of 3C~105, we also combined the spectra of the three
observations spanning over 2 days (Table~\ref{tab:obs}). 
Considering the lower energy resolution of \textit{NuSTAR}/FPMs than \textit{Suzaku}/XIS
and the absence of sharp spectral features above 8 keV we use 
(see Section~\ref{sec:x-ray}), we combined the spectral bins to contain at least 30
photons per bin (cf. 25 photons per bin for \textit{Suzaku}/XIS), in order
to increase the sensitivities for continuum determination.

\section{X-RAY SPECTRAL ANALYSIS} \label{sec:x-ray}

We simultaneously fit the broadband spectra of each object taken with
\textit{Suzaku}/XIS-FI (0.5--10~keV),
\textit{Suzaku}/XIS-BI (0.5--9~keV), 
\textit{Suzaku}/HXD-PIN (16--40~keV), and
\textit{NuSTAR}/FPMs (8--66~keV; for the sources for which it was possible to have the widest spectral coverage)
\footnote{We do not include the HXD-PIN spectra for PKS~0326--288 and 3C~452 in the fitting, considering the poor signal-to-noise ratios and possible contamination by nearby sources, respectively.}.
Conservatively, we decide not to use the \textit{NuSTAR} data below 8 keV,
considering the
possible cross-calibration uncertainties in the energy-dependent
effective area with other instruments at energies below 6--8 keV (see
e.g., Appendix 3 of \citealp{Diez23}).
Spectral fitting is performed on XSPEC \citep{Arnaud96} v12.12.1 based on
$\chi^2$-statistics.

\subsection{Spectral Model} \label{subsec:model}
For consistency with previous works on radio-quiet AGNs utilizing
the XCLUMPY model (e.g., \citealt{Tanimoto20, Ogawa21}; Inaba et al, \emph{in prep}),
we adopt basically the same spectral model as used in these papers. 
Our model represents a typical X-ray spectrum of an obscured AGN, 
consisting of a direct power-law component absorbed by material in the line of sight, 
reflection components from the torus, and an unabsorbed scattered
component by optically thin matter (most likely by an ionized gas in the polar region).
In many cases, we also add optically-thin thermal emission from
the host galaxy that is seen low energies below $\sim$2~keV.
The model is expressed as follows in XSPEC terminology:

\begin{eqnarray}
\label{equation:model}
\mathrm{Model} &=& \textsf{const1*phabs} \nonumber\\
&*& (\textsf{zphabs*cabs*const2*zcutoffpl} \nonumber\\
&+& \textsf{atable\{xclumpy\_RC.fits\}} \nonumber\\
&+& \textsf{atable\{xclumpy\_RL.fits\}} \nonumber\\
&+& \textsf{const3*zcutoffpl } \nonumber\\
&+& \textsf{apec}) \nonumber
\end{eqnarray}

\begin{enumerate}
\renewcommand{\labelenumi}{(\arabic{enumi})}
\item The \textsf{const1} term represents an instrumental cross-calibration constant among different detectors. We set \textsf{const1} of \textit{NuSTAR}/FPM
to 1 as a reference, and fix those of \textit{Suzaku}/XIS-FI and \textit{Suzaku}/HXD at 0.95 and 1.12, respectively, based on the results obtained by \cite{Madsen17}.
We allow that of \textit{Suzaku}/XIS-BI to be free.
The \textsf{phabs} term represents the Galactic absorption, whose
hydrogen column density ($N^{\mathrm{Gal}}_{\mathrm{H}}$) is fixed
at the total value of $HI$ and $H_{2}$ based on
https://www.swift.ac.uk/analysis/nhtot/ \citep{Willingale13}.

\item The \textsf{zcutoffpl} term, a power law with a high energy cutoff,
represents the direct component. We fix the cutoff energy
at a typical value ($E_{\mathrm{cut}} = 370$ keV: \citealt{Ricci18}), since it would be difficult to constrain from our data.
The \textsf{zphabs*cabs} term accounts for the photoelectric absorption and Compton scattering by line-of-sight material, 
whose hydrogen column density is determined by equation~\ref{N_HLOS}.
The \textsf{const2} term is a constant to deal with time variability between the \textit{NuSTAR} and \textit{Suzaku} observations.
We fix that of \textit{NuSTAR} at unity as a reference, and allow that of \textit{Suzaku} free.
Here we implicitly assume that the spectral shape is constant between the different observation epochs. 
Indeed, by separately analyzing the \textit{Suzaku} and \textit{NuSTAR} spectra at energies above 8 keV 
(commonly covered by the two satellites) with a simple absorbed power-law model, we confirmed that the 
spectral parameters did not vary over the statistical errors. It is also supported by the absence of any 
systematic residuals in the simultaneous fit of the \textit{Suzaku} and \textit{NuSTAR} spectra (Figure~\ref{fig:plots}).

\item The two tables of XCLUMPY, \textsf{atable\{xclumpy\_RC.fits\}} and \textsf{atable\{xclumpy\_RL.fits\}}, which
correspond to the reflection continuum and emission line components, respectively. 
The free parameters of XCLUMPY are the hydrogen column density along the equatorial plane $(N^{\rm Equ}_{\rm H})$, 
the torus angular width $(\sigma)$, and the inclination angle $(i)$.
Considering that all targets show large line-of-sight absorption, we
limit the inclination to a range between $60^{\circ}$ and $87^{\circ}$
(the hard limit in the table models).  
For 3C~105, we fix it at $i=70^{\circ}$, since it cannot be constrained by our data. 
The photon index, normalization, and cutoff energy are linked to the values of the direct component. 
The hydrogen column density along the line of sight ($N^{\rm LOS}_{\rm H}$) in (2)
is calculated from the torus parameters as follows:
\begin{eqnarray} \label{N_HLOS}
   N^{\rm LOS}_{\rm H}\left(\theta\right) = N^{\rm Equ}_{\rm H} \exp\left(-\left(\frac{90^{\circ}-i}{\sigma}\right)^2\right).
\end{eqnarray}
Considering the large size of the torus ($\sim$pc), 
we assume that the flux of the torus reflection component remained constant among the different observation epochs.

\item The \textsf{const3} factor gives the scattering fraction ($f_{\rm scat} $).
The parameters of \textsf{zcutoffpl} are linked to those of the direct component. 
We limit $f_{\rm scat} $ within a range of 0.0--10~\%. 

\item The \textsf{apec} term represents optically-thin thermal emission from the host galaxy, 
which is added when required by the data.
  
\end{enumerate}

\begin{figure*}
\includegraphics[width=7.7cm]{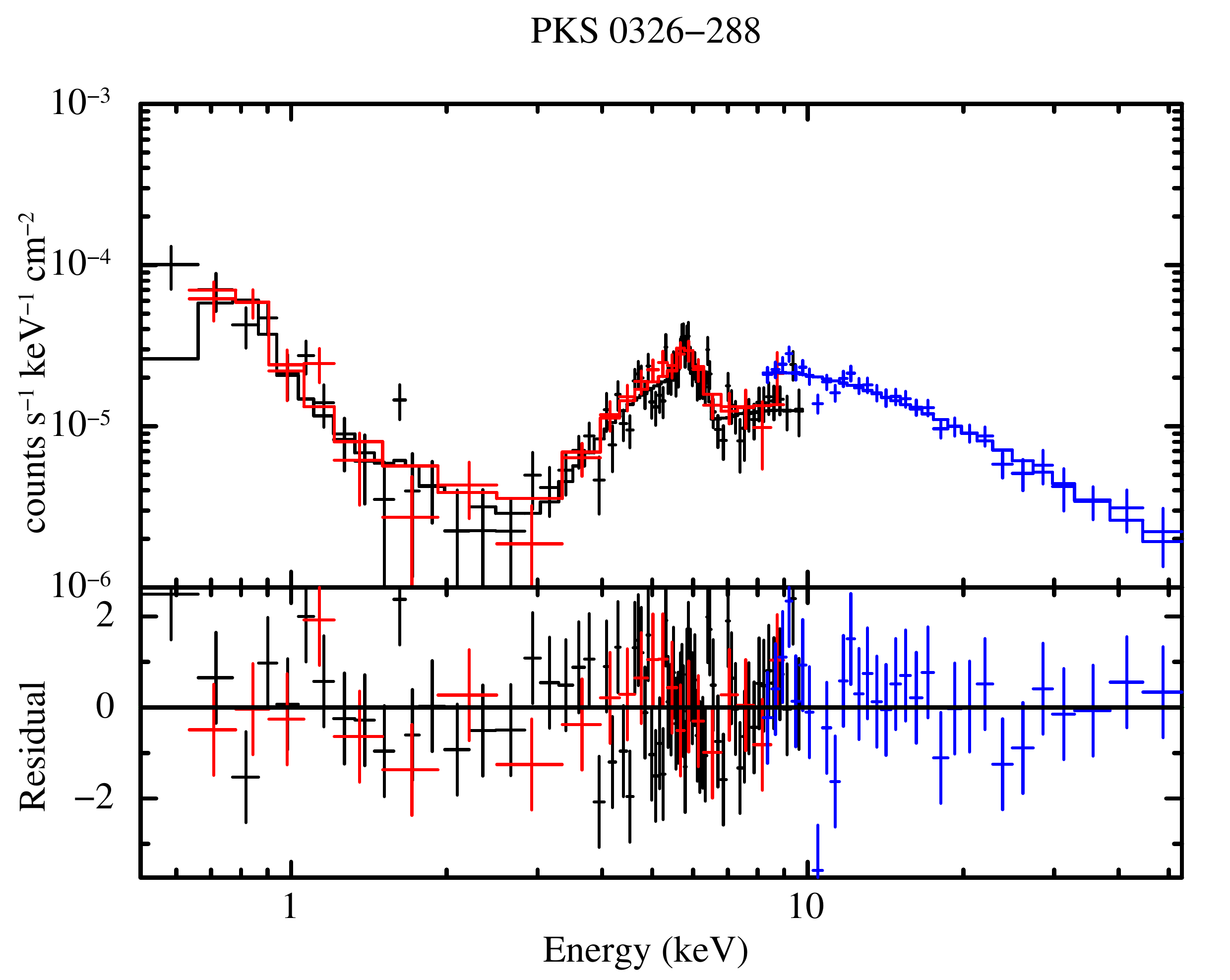}
\includegraphics[width=7.7cm]{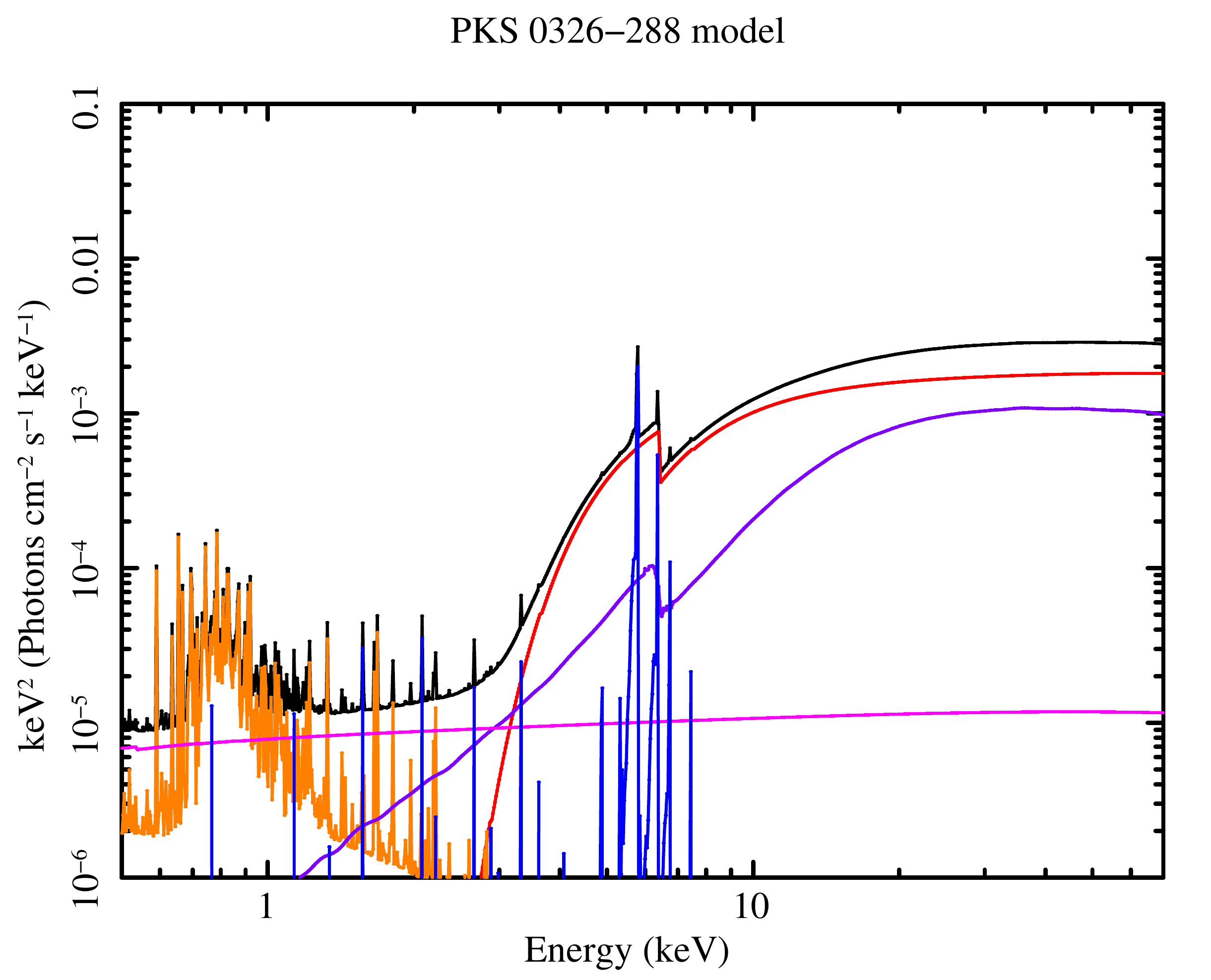}
\includegraphics[width=7.7cm]{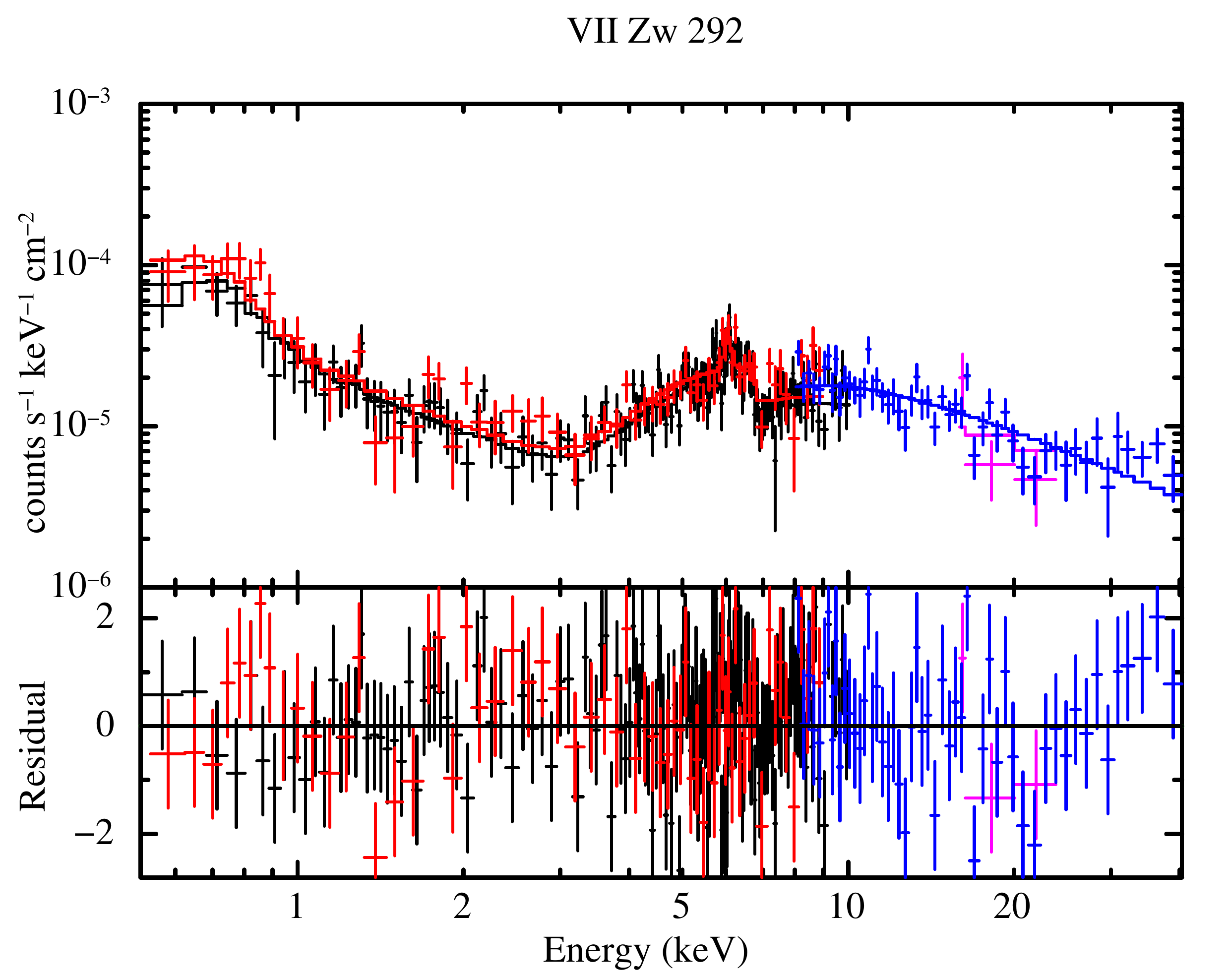}
\includegraphics[width=7.7cm]{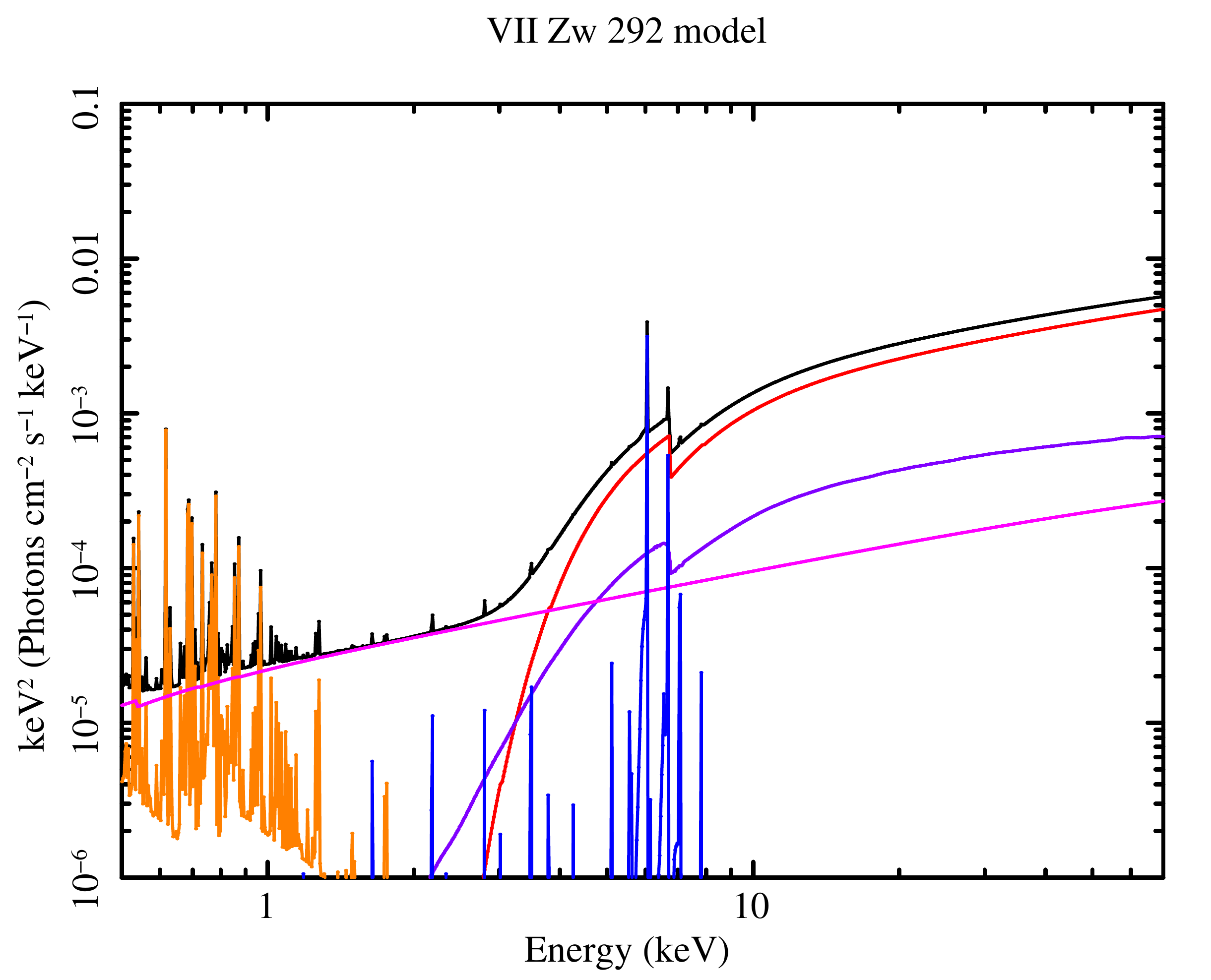}
\includegraphics[width=7.7cm]{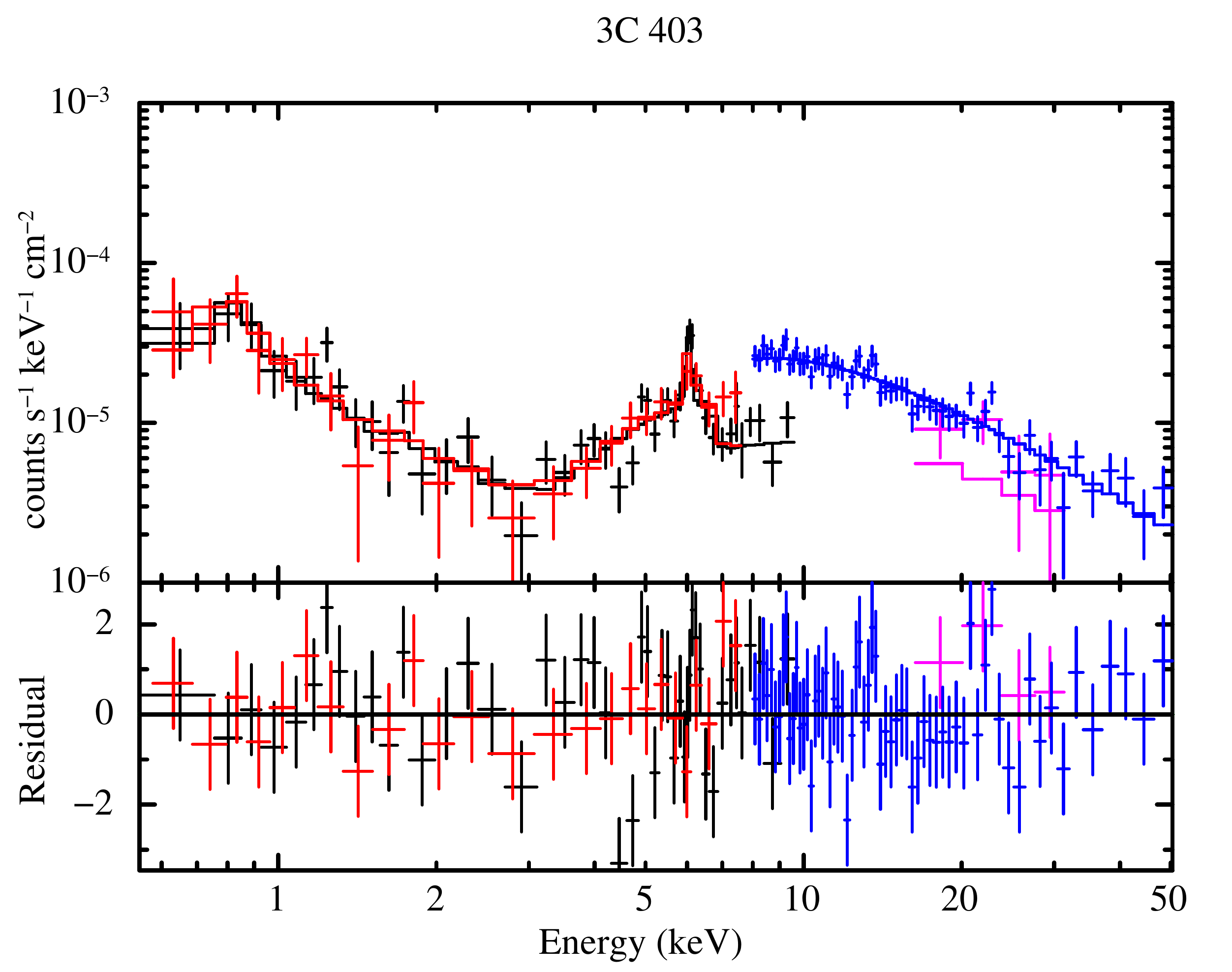}
\includegraphics[width=7.7cm]{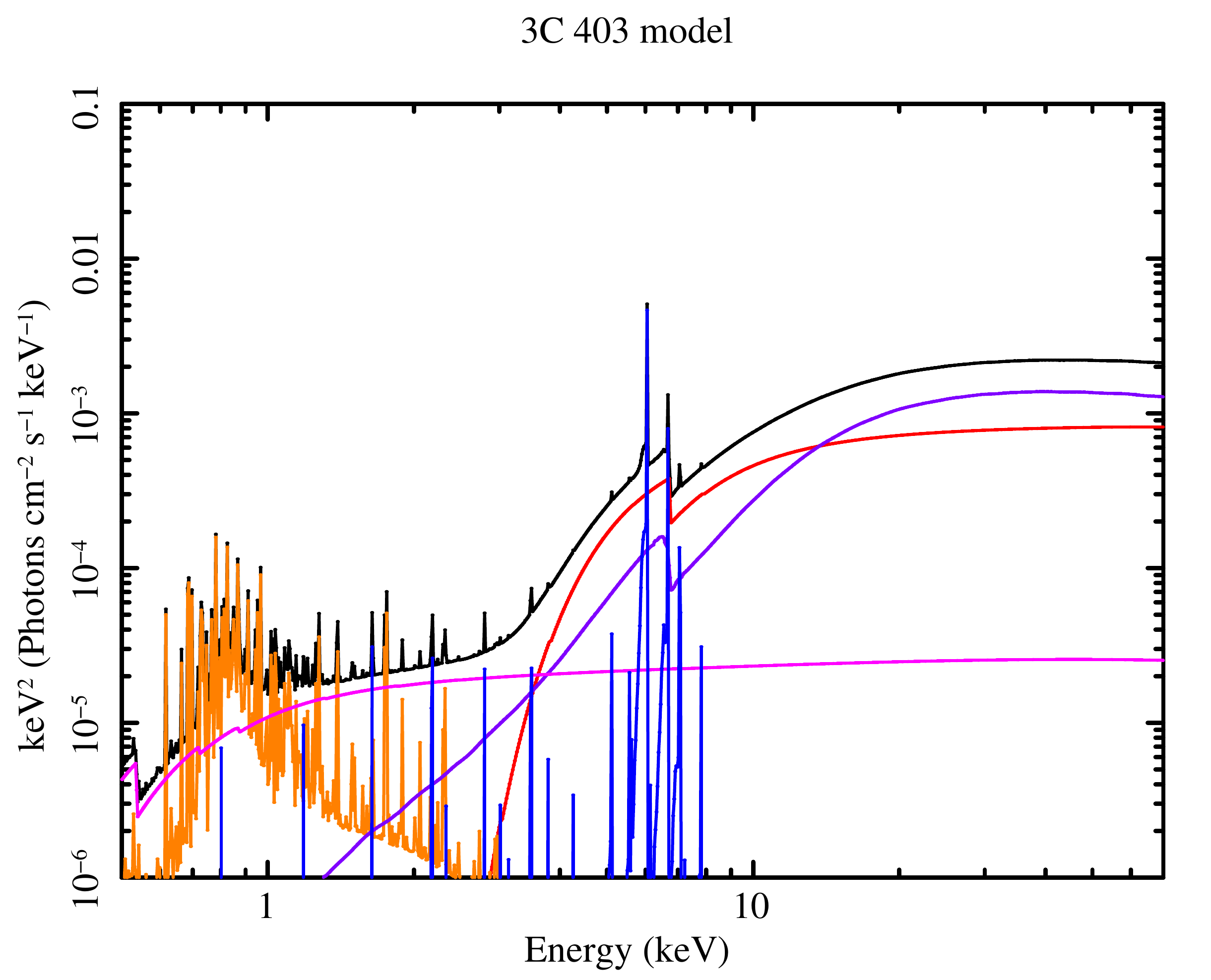}
\caption{Left: (upper panels) Observed X-ray spectra folded with the energy responses (but corrected for effective area): 
\textit{Suzaku}/XIS-FI (black crosses), \textit{Suzaku}/XIS-BI (red crosses), \textit{Suzaku}/HXD-PIN (magenta crosses), and
\textit{NuSTAR}/FPMs (blue crosses).
The best-fit models are overplotted by the lines.
(lower panels) Fitting residuals in units of 1$\sigma$ error. 
Right: The best-fit models in units of $E\times I_E$ (where $I_E$ is the energy flux at the energy $E$). 
The solid lines show the total (black), direct component (red), reflection continuum from the torus (purple), 
emission lines from the torus (blue), scattered component (magenta), and soft excess (orange).
}
\label{fig:plots}
\end{figure*}

\begin{figure*}
\includegraphics[width=7.7cm]{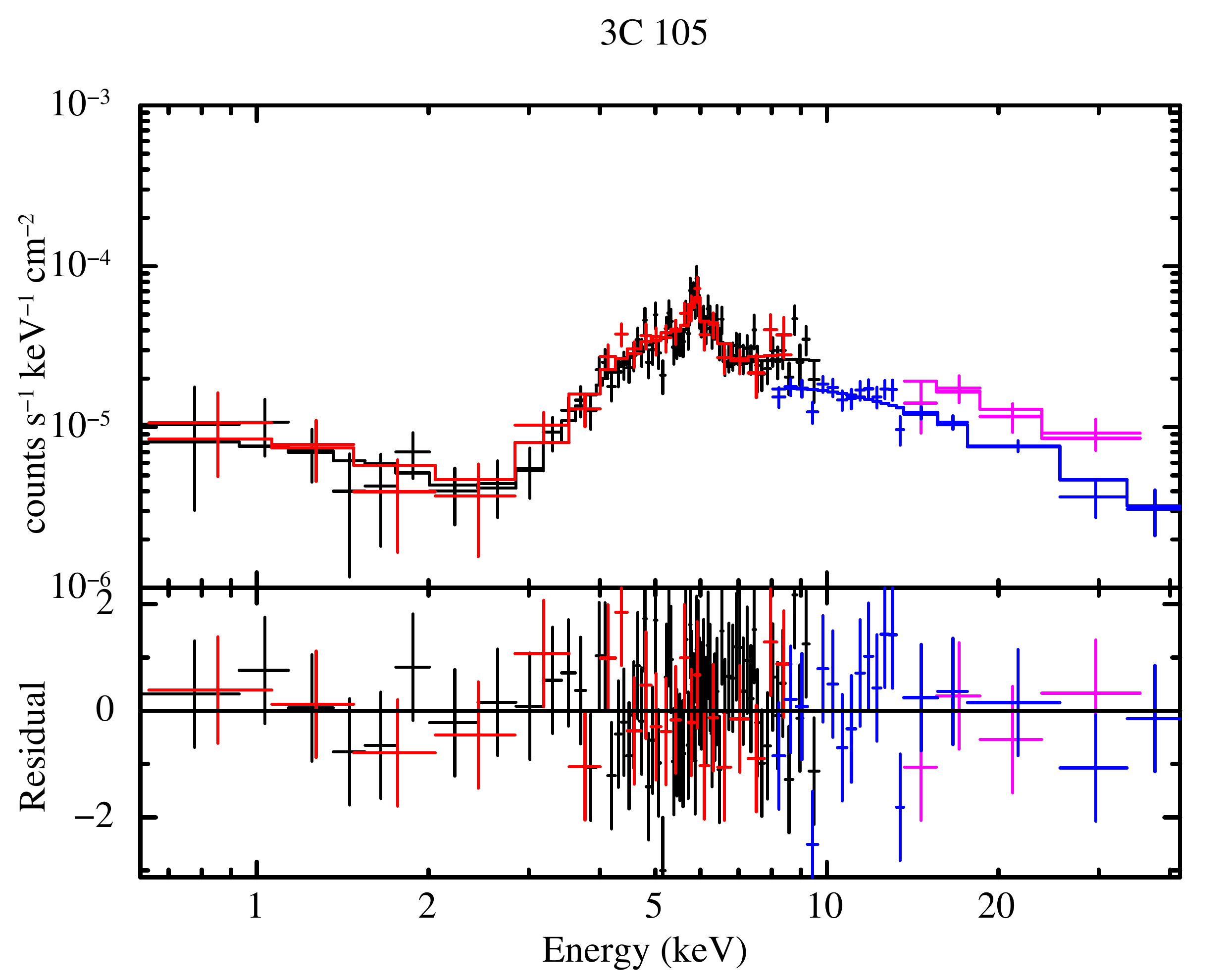}
\includegraphics[width=7.7cm]{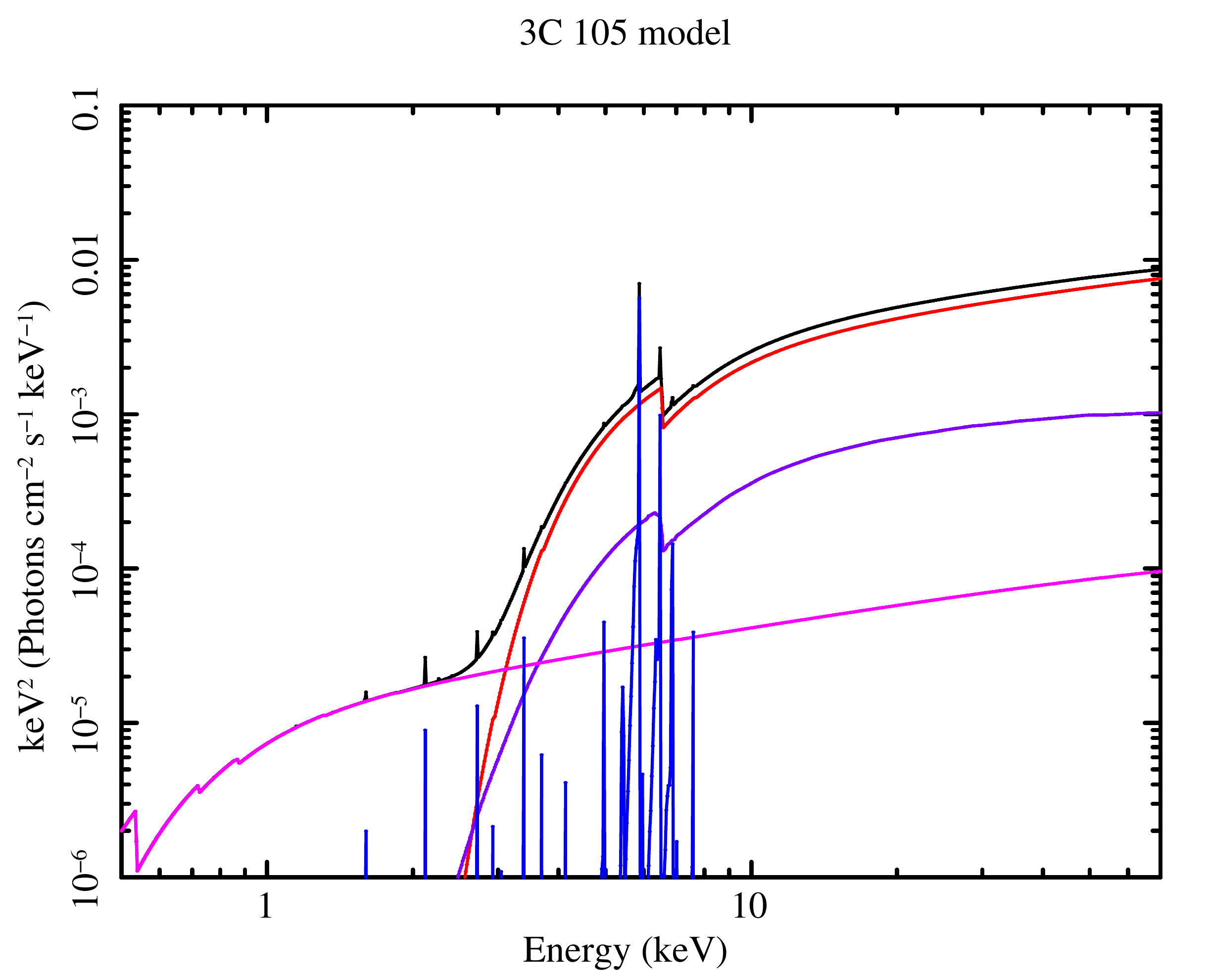}
\includegraphics[width=7.7cm]{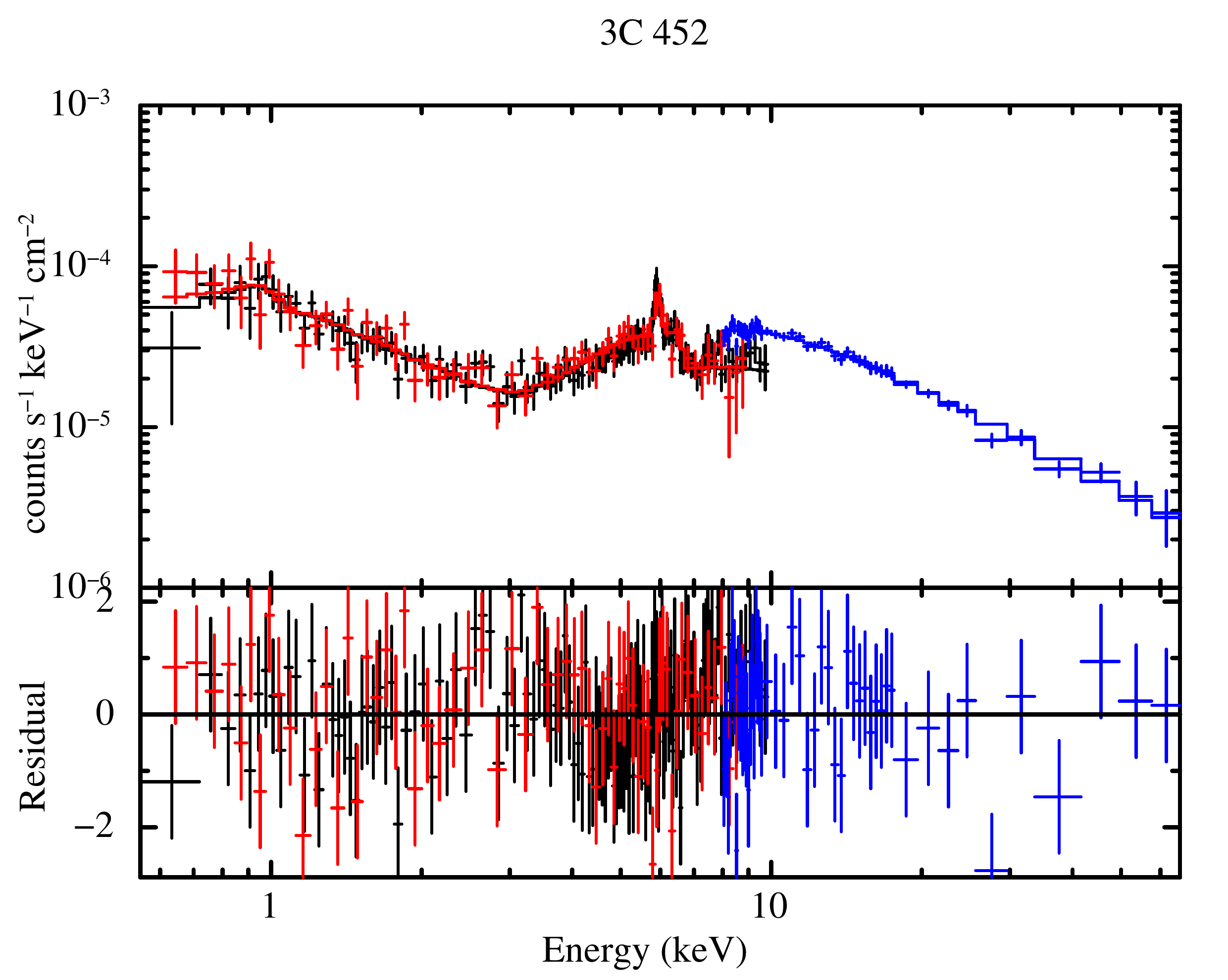}
\includegraphics[width=7.7cm]{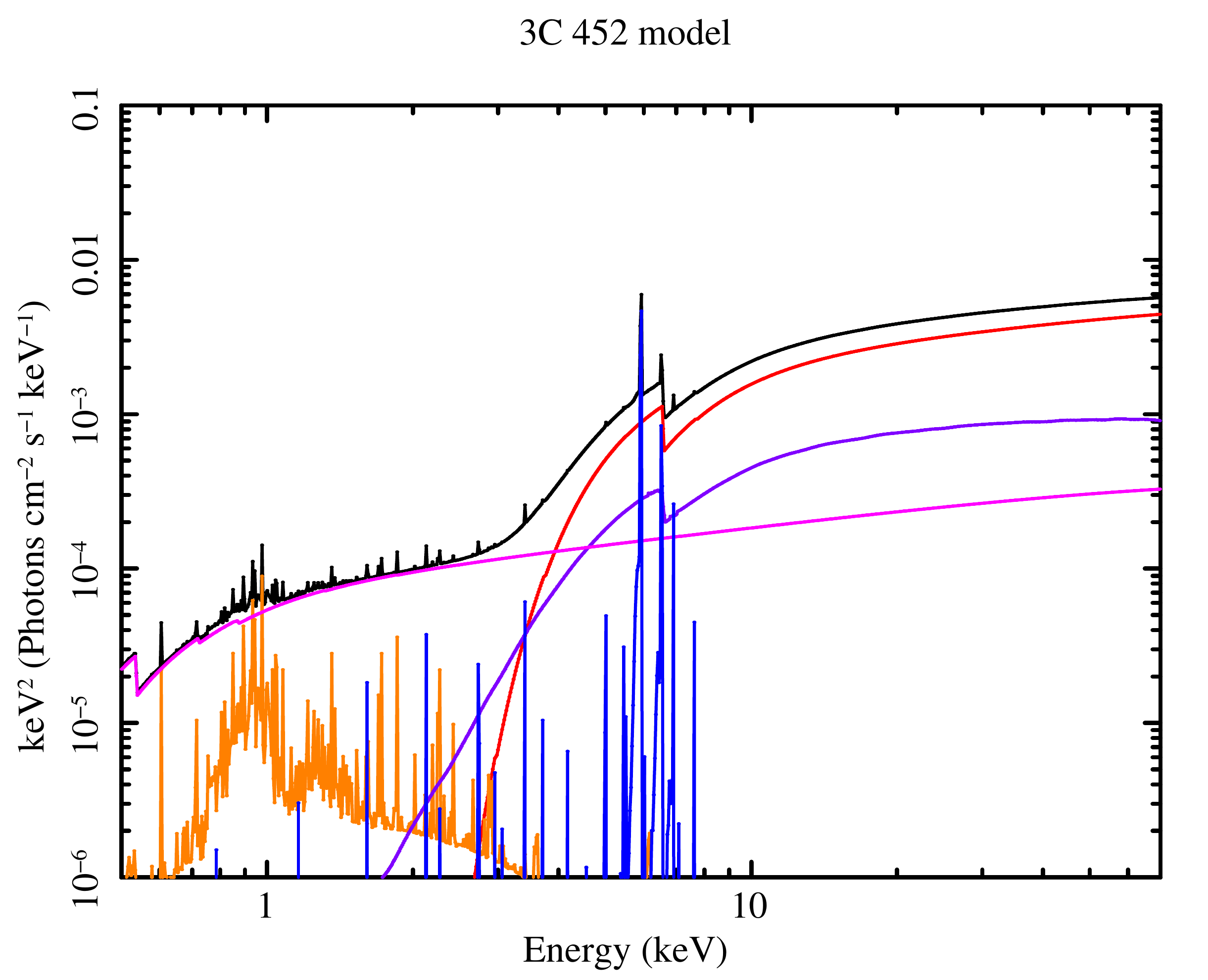}
\includegraphics[width=7.7cm]{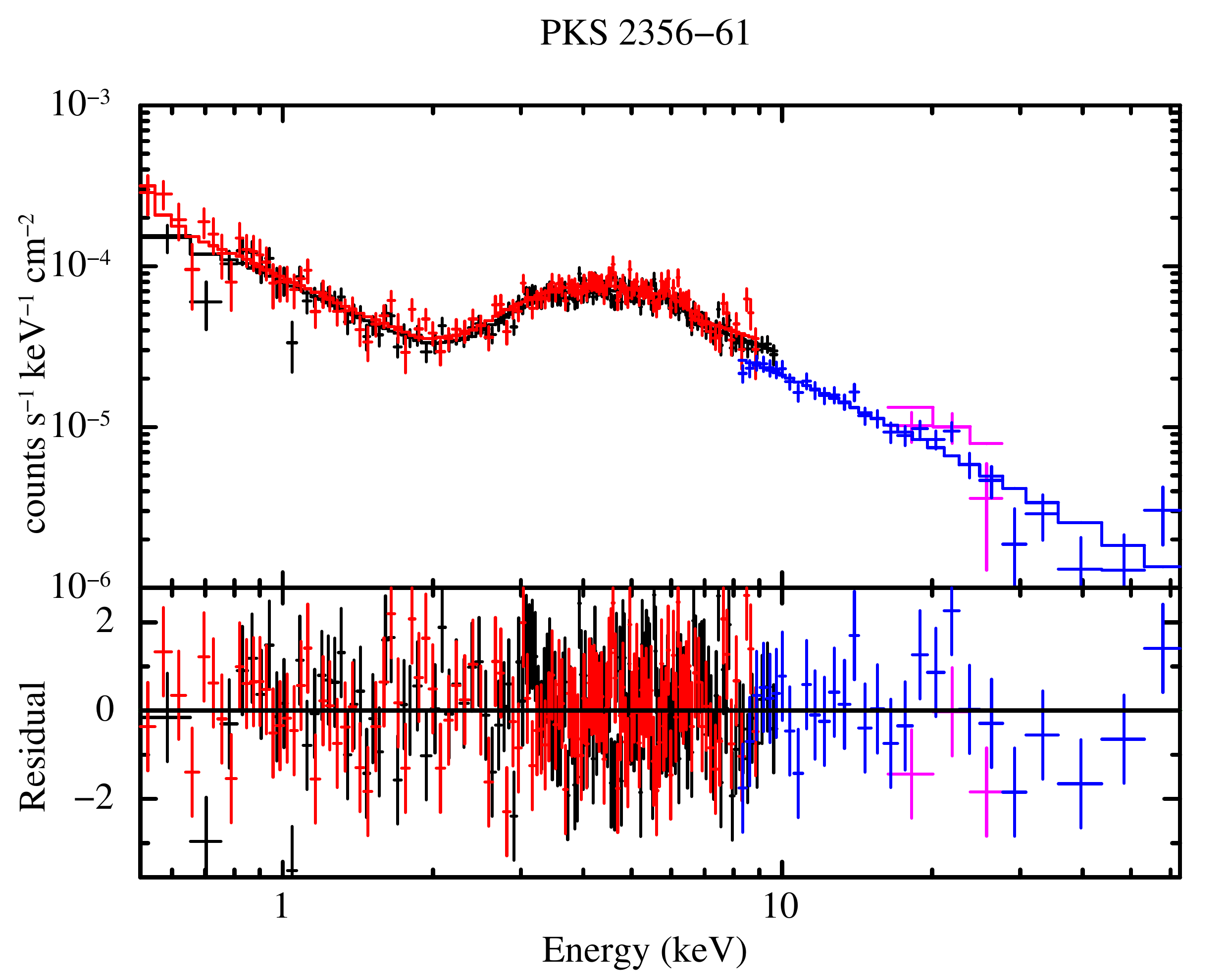}
\includegraphics[width=7.7cm]{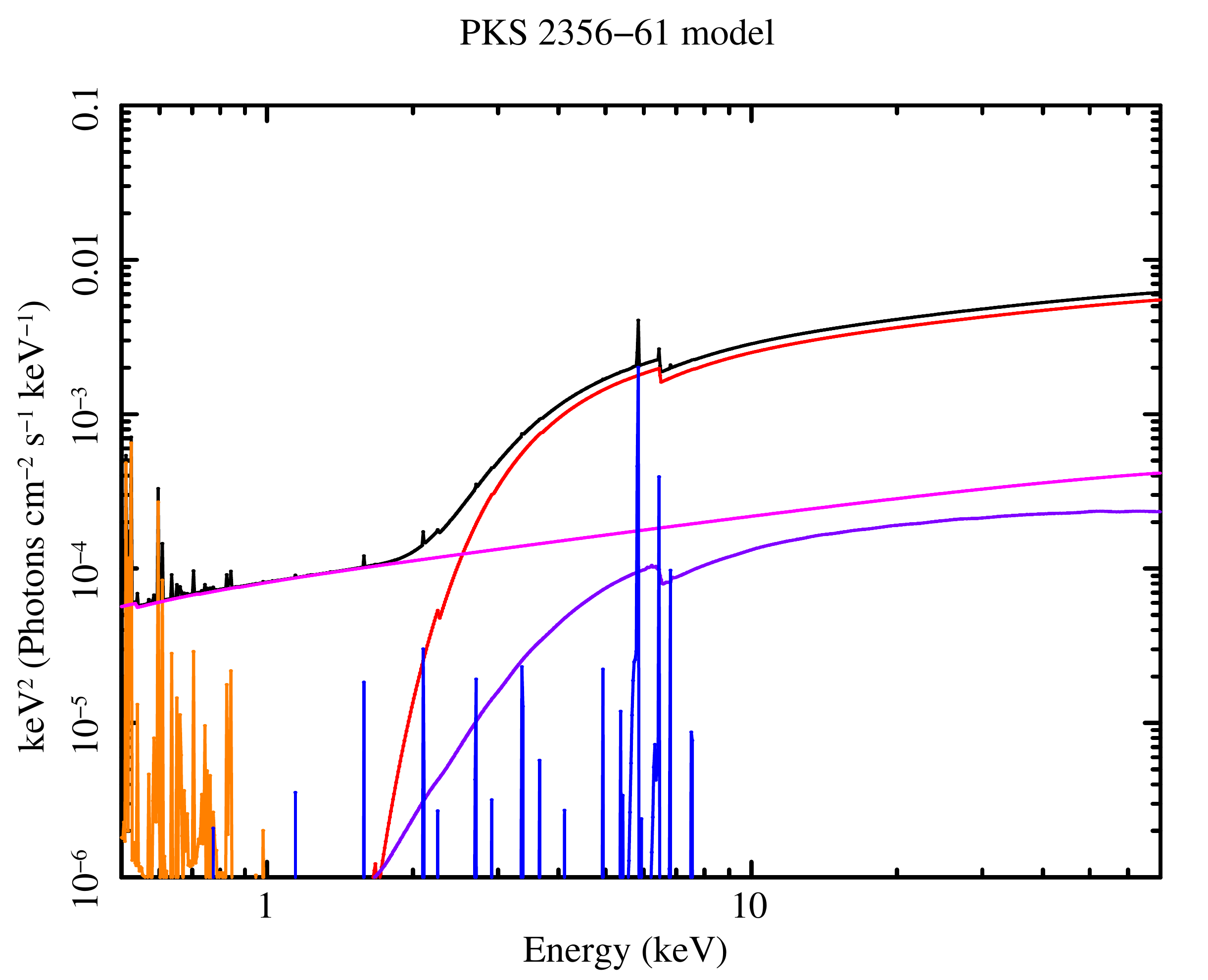}
\setcounter{figure}{1}
\caption{Continued.}
\end{figure*}

\subsection{Results} \label{subsec:result}

We find that this model can fairly reproduce the observed broadband
spectra of all objects ($\chi^2_{\mathrm{red}}\equiv\chi^2/{\rm d.o.f}< 1.2$).
The best-fit parameters are summarized in Table~\ref{tab:best}.  
The observed spectra folded with the energy responses (corrected for
effective area) are plotted in the left column of
Figure~\ref{fig:plots}, where the best-fit models are overplotted
together with the fitting residuals (lower panels). In the right
column of Figure~\ref{fig:plots}, we plot the multi-component best-fit
models in units of $E\times I(E)$
(where $I(E)$ is the energy flux density at the energy $E$).
Table~\ref{tab:critical} lists
the intrinsic (de-absorbed) 2--10 keV 
luminosities of the direct components obtained from our spectral fitting,
together with the black hole masses and Eddington ratios taken from \citet{Koss22}
\footnote{Throughout the paper, we adopt the Eddington ratio values in \citep{Koss22}, which were obtained from the 105-month averaged Swift/BAT data, in order to minimize the effect of time variability.}.

\begin{table*}
\caption{Best Fit Parameters.\label{tab:best}
\newline
(1) Galaxy name. 
(2) Logarithmic observed flux in ergs cm$^{-2}$ s$^{-1}$, in the range of 2--10 keV (\textit{Suzaku}/XIS-FI).
(3) Photon index of the direct component, and 
the normalization in 10$^{-3}$~photons~keV$^{-1}$~cm$^{-2}$~s$^{-1}$ at 1~keV. 
(4) Temperature of the \textsf{apec} model in keV and 
its normalization of the \textsf{apec} model in 10$^{-19}/4\pi [D_{\mathrm{A}}(1+z)]^2 \int n_{\mathrm{e}} n_{\mathrm{H}}dV$, 
where $D_{\mathrm{A}}$ is the angular diameter distance to the source in cm, 
$n_{\mathrm{e}}$ and $n_{\mathrm{H}}$ are the electron and hydrogen densities in cm$^{-3}$, respectively.
(5) Hydrogen column density along the equatorial plane in 10$^{24}$~cm$^{-2}$,
torus angular width in degree,
inclination angle of the observer in degree,
and hydrogen column density along the line of sight in 10$^{22}$~cm$^{-2}$.
(6) Time variability constant (relative normalization of \textit{Suzaku}/XIS to \textit{NuSTAR}/FPMs) 
and instrumental cross-calibration constant of \textit{Suzaku}/XIS-BI against \textit{Suzaku}/XIS-FI.
(7) Scattering fraction in \%.
\newline
$\dagger$ - The parameter has reached its limit of its allowed range.
$\ddagger$ - The parameter has been fixed at $70^{\circ}$.
}
\begin{tabular}{llllllll}
\hline
(1)&(2)&(3)&(4)&(5)&(6)&(7)&\\
Object            &$\log F_{2-10 \rm keV}$ &\textsf{zcutoffpl}  &\textsf{apec}     &\textsf{XCLUMPY}&
 \textsf{const2}    &\textsf{const3}      &  \\
                  &&$\Gamma$            &$kT$              &$N_{\mathrm{H}}^{\mathrm{Equ}}$ &
$C_{\mathrm{TIME}}$& $f_{\mathrm{scat}} $    &  $\chi^2/$d.o.f.\\
                  &&$N_{\mathrm{pow}}$   &$N_{\mathrm{apec}}$&
$\sigma $         &    $C_{\mathrm{BI}}$ &   &\\
&&& & $i$ &&&\\
&&& & $N_{\mathrm{H}}^{\mathrm{LOS}}$&&&\\
\hline

PKS~0326--288	&-12.0	&$1.74_{-0.14}^{+0.13}$	&$0.79_{-0.17}^{+0.15}$	&$1.2_{-0.7}^{+8.8\dagger}$	&$0.56_{-0.07}^{+0.07}$	&$0.38_{-0.15}^{+0.2}$	&$136.22/119$	\\
&&$3.0_{-1.07}^{+1.42}$	&$1.25_{-0.31}^{+0.32}$	&$10.0_{-0.0\dagger}^{+8.92}$	&$1.09_{-0.09}^{+0.1}$	&&\\
&&&&$80.69_{-16.21}^{+6.31\dagger}$	&&&\\
&&&&$49.7_{-6.79}^{+3.5}$	&&&\\
\hline
VII~Zw~292	&-11.95	&$1.36_{-0.13}^{+0.14}$	&$0.29_{-0.05}^{+0.11}$	&$0.42_{-0.05}^{+0.98}$	&$0.74_{-0.09}^{+0.11}$	&$3.36_{-1.02}^{+1.44}$	&$300.25/264$	\\
&&$0.79_{-0.27}^{+0.43}$	&$2.28_{-0.92}^{+0.96}$	&$26.08_{-14.49}^{+14.54}$	&$1.11_{-0.07}^{+0.07}$	&&\\
&&&&$87.0_{-19.26}^{+0.0\dagger}$	&&&\\
&&&&$41.14_{-2.49}^{+2.71}$	&&&\\
\hline
3C~403	&-12.17	&$1.87_{-0.24}^{+0.19}$	&$0.73_{-0.19}^{+0.15}$	&$5.62_{-4.85}^{+4.38\dagger}$	&$0.22_{-0.05}^{+0.04}$	&$0.51_{-0.24}^{+0.48}$	&$152.23/130$	\\
&&$4.08_{-2.14}^{+3.86}$	&$1.85_{-0.57}^{+0.61}$	&$14.46_{-4.46\dagger}^{+4.43}$	&$1.03_{-0.12}^{+0.13}$	&&\\
&&&&$66.89_{-6.89\dagger}^{+10.06}$	&&&\\
&&&&$43.65_{-6.71}^{+7.1}$	&&&\\
\hline
3C~105	&-11.68	&$1.48_{-0.12}^{+0.12}$	&$\cdots$	&$0.41_{-0.04}^{+0.19}$	&$1.72_{-0.18}^{+0.18}$	&$1.7_{-0.55}^{+0.47}$	&$103.17/113$	\\
&&$0.9_{-0.22}^{+0.39}$	&$\cdots$	&$89.59_{-57.58}^{+0.41\dagger}$	&$1.07_{-0.09}^{+0.09}$	&&\\
&&&&$70.0\ddagger$	&&&\\
&&&&$39.27_{-3.58}^{+2.97}$	&&&\\
\hline
3C~452	&-11.69	&$1.61_{-0.07}^{+0.06}$	&$1.22_{-0.78}^{+0.7}$	&$0.44_{-0.04}^{+1.11}$	&$0.5_{-0.04}^{+0.05}$	&$2.79_{-0.45}^{+0.68}$	&$192.64/230$	\\
&&$3.32_{-0.76}^{+0.79}$	&$1.34_{-1.03}^{+2.17}$	&$22.11_{-9.39}^{+6.04}$	&$1.02_{-0.05}^{+0.05}$	&&	\\
&&&&$86.63_{-15.56}^{+0.37\dagger}$	&&&\\
&&&&$43.39_{-5.78}^{+2.24}$	&&&\\
\hline
PKS~2356--61	&-11.45	&$1.58_{-0.01}^{+0.05}$	&$0.16_{-0.15}^{+0.05}$	&$0.27_{-0.14}^{+0.86}$	&$1.44_{-0.09}^{+0.06}$	&$10.0_{-1.27}^{+0.0\dagger}$	&$339.1/306$	\\
&&$1.03_{-0.05}^{+0.14}$	&$2.49_{-1.14}^{+3.92}$	&$24.14_{-6.56}^{+65.86\dagger}$	&$1.07_{-0.03}^{+0.03}$	&&	\\
&&&&$70.09_{-5.05}^{+16.91\dagger}$	&&&\\
&&&&$13.64_{-0.34}^{+0.56}$	&&&\\
\hline

\end{tabular}
\end{table*}
\begin{table*}
  \caption{AGN key parameters.\label{tab:critical}
\newline
(1) Galaxy name.
(2) Logarithmic black hole mass in $M_\odot$ \citep{Koss22}. 
(3) Logarithmic Eddington ratio \citep{Koss22} based on the 105-month averaged Swift/BAT luminosity.
(4) Logarithmic intrinsic 2--10 keV luminosity of the direct component in erg s$^{-1}$ derived from our spectral analysis.
(5) Torus covering fraction.
}
\begin{tabular}{lllll}
\hline
(1)&(2)&(3)&(4)&(5)\\
Object                          & $\log M_{\mathrm{BH}}/M_\odot$      
&$\log \lambda _{\mathrm{Edd}}$ 
&$\log L_{2-10 \rm keV}$ &$C_\mathrm{T}$\\
\hline
PKS~0326--288	&8.48		&-1.14 &44.49	&$0.37_{-0.03}^{+0.27}$	\\
VII~Zw~292	&8.74	&-1.84	&43.6	&$0.77_{-0.32}^{+0.21}$	\\
3C~403	&8.83	&-1.59	&43.97	&$0.59_{-0.23}^{+0.14}$	\\
3C~105	&8.59	&-1.07	&43.94	&$1.0_{-0.1}^{+0.0}$	\\
3C~452	&8.89	&-1.46	&44.33	&$0.68_{-0.23}^{+0.16}$	\\
PKS~2356--61	&8.55  &-1.36	&43.99	&$0.69_{-0.14}^{+0.31}$	\\
\hline
\end{tabular}
\end{table*}

\section{DISCUSSION} \label{sec:discussion}

\subsection{Properties of Local Obscured Radio Galaxies} \label{subsec:result_summary}

For our study, we have constructed a new sample of nearby X-ray obscured
NLRGs consisting of the seven most radio-loud AGNs
with absorption of $\log (N^{\rm LOS}_{\rm H}/\rm cm^{-2}) > 23$
in the \textit{Swift}/BAT 70-month AGN (non-blazar) catalog. 
Most of them are classified as Fanaroff-Riley (FR) II RGs
(Table~\ref{tab:obj}), which have stronger jet power than FR I RGs (\citealt{Fanaroff&Riley74}).
We have adopted the radio-to-X-ray radio loudness
parameter $\rm R \equiv \log (\nu f_{\nu(1.4 \rm GHz)}/f_{14-150 \rm keV}) > -2.8$, 
which is free from AGN obscuration and contamination
by the host galaxy emission; these are advantages to obtain a
``clean'' radio-loud AGN sample over radio loudness parameters defined
between radio and optical fluxes (\citealt{Terashima&Wilson03}).
Our sample is
selected with well-defined
criteria (Section~\ref{sec:sample}) except for the availability of high-quality hard
band data taken with \textit{NuSTAR},
and hence can be regarded as a representative sample of
local obscured RGs with strong jets.

As mentioned in Section~\ref{sec:sample}, most of
our sample (except Centaurus A) has limited ranges of
the bolometric luminosity ($45<\log L_{\rm bol}/({\rm erg s}^{-1}) <46$) and 
Eddington ratio ($-2<\log \lambda_{\rm Edd}<-1$).
Due to the small size of our sample, it is not clear whether the most powerful jets in AGNs can be selectively produced at specific conditions on luminosity or Eddington ratio; 
quantitative comparison of luminosity or Eddington ratio distribution between radio-loud and radio-quiet AGNs
is out of the scope of this paper and is left for a future work.
Nevertheless, since these values of $\log L_{\rm bol}$ and $\lambda_{\rm Edd}$ are typical of \textit{Swift}/BAT AGNs (mostly radio-quiet ones),
it is suggested that the
launching mechanisms of jets are not controlled solely by the mass
accretion rate in the accretion disk (luminosity)
or that normalized by the black hole mass (Eddington ratio).
This is consistent with the presence of two distinct sequences in the
Eddington ratio versus radio loudness plane found for radio- or
optically-selected AGNs (\citealt{Sikora07}).

\subsection{Torus Covering Fraction as a Function of Eddington Ratio} \label{subsec:Torus_Covering_Fraction}

We have 
analyzed the currently available highest-quality broadband
X-ray spectra of six RGs obtained with \textit{Suzaku} and
\textit{NuSTAR} in the 0.5--66~keV band. 
To study the X-ray reprocessed radiation from the torus, 
we have employed one of the latest X-ray clumpy models, XCLUMPY \citep{Tanimoto19}.
In this subsection, we discuss the torus geometry of our RGs in terms
of the covering fraction $C_\mathrm{T}$, defined as the solid angle of the torus matter with column densities of $\log N_{\rm H}/{\rm{cm}^{-2}}
>22$ normalized by $4\pi$.

By measuring absorbed AGN fractions in the \textit{Swift}/BAT AGN sample, which
is dominated by radio-quiet AGNs, \citet{Ricci17Nature} find that the
torus covering fraction is determined mainly by the Eddington ratio, not by
the luminosity of the AGN
(see also \cite{Ricci22} for an updated view).
Since $C_\mathrm{T}$ shows a rapid drop at $\log \lambda_{\rm Edd}\gtrsim -1.5$ , they conclude that
obscuring matter in an AGN is located within the sphere of influence by
the SMBH and the torus is shaped by radiation pressure of the AGN on dusty gas.
Utilizing the best-fit torus parameters of XCLUMPY, it is also
possible to calculate $C_{\mathrm{T}}$ of an individual object
as follows \citep{Ogawa21}:
\begin{eqnarray} \label{C_T}
   C_{\rm T} = \rm{sin}\Biggl(\sigma\sqrt{\rm{ln}\biggl(\frac{N^{\rm{Equ}}_{\rm{H}}}{10^{22}\ \rm{cm}^{-2}}\biggr)}\Biggr).
\end{eqnarray}
\citet{Ogawa21} and Inaba et al, (\emph{in prep}) confirm that the covering fractions of
radio-quiet AGNs determined through
X-ray spectroscopy follow well the \citet{Ricci17Nature} relation between
$C_\mathrm{T}$ and $\log \lambda_{\rm Edd}$.

\begin{figure}
\includegraphics[width=8.5cm]{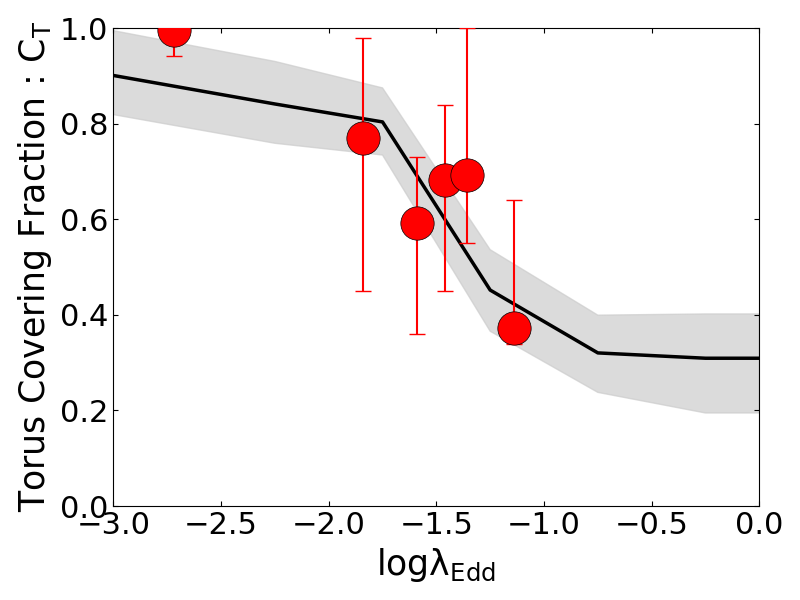}
\caption{The torus covering fraction ($C_\mathrm{T}$)
with $\log N_{\rm H}/{\rm{cm}^{-2}} >22$
as a function of Eddington ratio ($\lambda_{\mathrm{Edd}}$).
The red circles represent our RG sample including Centaurus A (left-most, \citealt{Ogawa21}).
The relation obtained by
\citet{Ricci17Nature} for the \textit{Swift}/BAT 70 month AGN
sample is indicated by the black line (best-fit) and gray
hatched area (1$\sigma$ uncertainties).
3C~105, whose $C_\mathrm{T}$ is unconstrained by our data,
is excluded in this plot.
}
\label{fig:Ricci_diagram}
\end{figure}

Table~\ref{tab:critical} lists the $C_\mathrm{T}$ values for our RG sample.
The result for Centaurus A is taken after \cite{Ogawa21}.
Figure~\ref{fig:Ricci_diagram} plots $C_\mathrm{T}$ as a function of
$\lambda_{\rm Edd}$ (taken from \citealp{Koss22});
here we exclude 3C~105 where the torus widths ($\sigma$) is almost unconstrained 
by our data even when the inclination is fixed. 
We estimate the errors of $C_\mathrm{T}$ by fully considering the 
parameter coupling among the torus parameters; we first make the
error contour map on the $N^{\rm{Equ}}_{\rm{H}}$ versus $\sigma$ plane and 
then find the error region of $C_\mathrm{T}$ by comparing the contours with constant $C_\mathrm{T}$ lines on the plane.

As noticed from the figure, our results for the individual objects 
are mostly consistent with the relation
by \citet{Ricci17Nature}. 
The mean value of the covering fractions for the five objects
at $-2 < \log \lambda_{\rm Edd} < -1$ is found to be
$\bar{C_\mathrm{T}}=0.62\pm0.08$.
This is similar to the
average of radio-quiet AGNs at the same range of $\log \lambda_{\rm Edd}$
obtained by Inaba et al, (\emph{in prep}), $\bar{C_\mathrm{T}}=0.62\pm0.02$.
Centaurus A shows $C_\mathrm{T}=0.995^{+0.005}_{-0.057}$ at $\log \lambda_{\rm Edd}=-2.7$,
which agrees with the \citet{Ricci17Nature} curve.

At even lower Eddington ratios, we refer to the Hitomi result of the
radio galaxy NGC 1275 (\citealt{Hitomi_Col18}), whose Eddington ratio
is estimated to be $\log \lambda_{\rm Edd} = -3.6$ (\citealt{Sikora07}).
On the basis of the narrow and weak iron-K$\alpha$ line, 
the authors concluded that the torus in NGC~1275 has only a small
solid angle. This trend is also seen in the total \textit{Swift}/BAT AGN sample
(\citealt{Ricci17Nature}), and is consistent with the spectral analysis
results of local low-luminosity AGNs
(e.g., \citealt{Kawamuro16} for AGNs with $<10^{42}$ erg s$^{-1}$ in the 14--195 keV band).
It may be explained by considering that, in a very low luminosity (or Eddington
ratio) AGN, gas in the torus or circumnuclear disk is depleted and
star formation activity is quenched; this prevents the torus from
being inflated and stops efficient mass transfer toward the SMBH, due
to the lack of turbulence by supernova feedback (\citealt{Kawakatu&Wada08}).

\subsection{Implications}  \label{subsec:Implications}

To summarize these results, we have shown that
the typical torus structure in RGs does not
show any distinct differences from radio-quiet AGNs at the same Eddington ratios, at least
in the range of $-3 < \log \lambda_{\rm Edd} < -1$ we have examined.
Although we have to bear in mind that the sample size is limited
to derive a conclusive argument, 
our work has two implications:

\begin{enumerate}
\item The torus structure is not an important factor that determines the presence or absence of powerful jets
at $-3 <\log \lambda_{\rm Edd} < -1$.
\item The jets have physically little effects on the torus structure.
\end{enumerate}

The first consequence can be understood if the launching mechanism of
powerful jets is critically related to 
physics in the innermost region of
the accretion disk (such as black hole spin and type of accretion flow)
and is not determined solely by the Eddington ratio as
discussed in Section~\ref{subsec:Torus_Covering_Fraction}. To check the second consequence
(i.e., if radiation from the jet core
could have significant impacts on the torus region), we refer to
\citet{Meyer11}, who systematically studied the synchrotron
peak luminosities ($\log L_{\rm peak}$) in radio-loud AGNs, including RGs and blazars.
Since our objects are viewed through the obscuring tori, we can
approximately use the observed jet luminosities as those actually
irradiating the tori without corrections for relativistic
beaming. Their figure~5 shows that FR II RGs have typically 
$\log (L_{\rm peak}$ erg s$^{-1})\sim 44.5$ with a core dominance parameter (the ratio of
the core luminosity to the total one including extended emission) of
$\log R_{\rm CE} \sim -1.5$. Thus, we can estimate the luminosities
from the jet core in typical FR II RGs to be $\sim10^{43}$ erg
s$^{-1}$, which is much smaller than the AGN bolometric luminosities
in our sample, $\sim10^{44-45}$ erg s$^{-1}$ (Figure~\ref{fig:Sample}). 
Hence, we infer that radiative feedback from the jet core can be neglected
compared with that from the accretion disk even in an AGN with powerful jets.

\section{CONCLUSION} \label{sec:conclusion}

We have reported the results of detailed broadband X-ray spectral
analysis of the seven
most radio-loud, X-ray obscured NLRGs in the
\textit{Swift}/BAT 70-month catalog, utilizing currently the highest-quality
data observed with \textit{Suzaku} and \textit{NuSTAR}.
The XCLUMPY model \citep{Tanimoto19} has been applied to the spectra,
which enables us to constrain the torus covering fraction of each
object. The main conclusions are summarized below.

\begin{itemize}

\item Most of our sample except Centaurus A cover a narrow range of AGN bolometric luminosities
  ($44<\log L_{\rm bol}$ erg s$^{-1}<45$) and Eddington ratios 
  ($-2 < \log \lambda_{\rm Edd} < -1$), 
  suggesting that the condition to launch powerful
  jets is not determined solely by the mass accretion rate or that 
  normalized by the Eddington ratio. 
  
\item The relation between the Eddington ratio and the torus covering
  fraction of the RGs follows the same trend for radio-quiet AGNs
  obtained by \citet{Ricci17Nature} based on a statistical analysis of
  obscured AGN fractions in a complete hard X-ray selected
  sample. This means that no distinct differences are found in the torus
  structure between AGNs with and without powerful jets.

\item Our results suggest that the torus structure is not an important factor that determines
  the presence of jets. This is probably because the launching
  mechanism of jets is determined by physics in the innermost region
  of the accretion disk and is unrelated to distant torus structure.
  Also, AGN jets have physically little effects on the torus. In fact,
  estimated radiation luminosities from the jet core to the torus
  region are much smaller than those from the accretion disk. 
  
\end{itemize}

\section*{Acknowledgements}
This work has been financially supported by JSPS KAKENHI
Grant Numbers 
20H01946 (Y.U.), 21J13894 (S.O.), 22J22795 (R.U.), and 
22K20391 and 23K13154 (S.Y.). S.Y. is grateful for support from RIKEN Special Postdoctral Researcher Program. 
C.R. acknowledges support from the Fondecyt Regular grant 1230345 
and ANID BASAL project FB210003.
We have made use of public data from Suzaku, which is obtained through
the Data ARchives and Transmission System (DARTS) supplied by the
Institute of Space and Astronautical Science (ISAS) at the Japan
Aerospace Exploration Agency (JAXA), and NuSTAR Data Analysis Software
(NuSTARDAS), which has been jointly developed by the Space Science
Data Center (SSDC; ASI, Italy) and the Jet Propulsion Laboratory
(JPL)/California Institute of Technology. We have also used NASA/IPAC
Extragalactic Database (NED), which is operated by JPL/California 
Institute of Technology, under contract with NASA.

\section*{Data Availability}
The datasets generated and/or analysed in this study are available
from the corresponding author on reasonable request.
\bibliographystyle{mnras}

\appendix

\section{COMPARISON OF SPECTRAL FITTING RESULTS WITH PREVIOUS
WORKS}\label{comparison}

In this section, we compare our spectral fitting results with those
reported in the literature for each object. Since various spectral
models were used in the literature, we focus on two basic continuum
parameters,
the photon index of the intrinsic power-law component ($\Gamma$) and
the line-of-sight hydrogen column density
($N_{\mathrm{H}}^{\mathrm{LOS}}$), to roughly check the
consistency with our results. \citet{Ricci17apj} (hereafter R17)
analyzed the broadband (0.3--150 keV) spectra of 838 AGNs including
our sample by utilizing the \textit{Swift}/BAT 70-month spectra and soft
X-ray data available then. They basically adopted the \textsc{pexrav} model
to represent the reflection component.
We note that \cite{Tombesi14} analyzed the same \textit{Suzaku} data
of all objects but VII~Zw~292. We do not refer to the
continuum parameters summarized in their Table~B2, however, because
only the 3.5--10.5 keV band spectra were utilized
to search for UFO features.

\subsection{PKS~0326--288}

The analysis results of the \textit{NuSTAR} data are
reported in this paper for the first time.
We obtain $\Gamma =
1.74_{-0.14}^{+0.13}
$ and $N_{\mathrm{H}}^{\mathrm{LOS}} =
49.7_{-6.79}^{+3.5}
\times 10^{22} \mathrm{cm}^{-2}$.
These values are consistent with those of R17 using the
\textit{Swift}/XRT and \textit{Swift}/BAT data ($\Gamma = 2.12_{-0.5
}^{+0.4 } $ and $N_{\mathrm{H}}^{\mathrm{LOS}} = 38.0_{-17.1}^{+62.0}
\times 10^{22} \mathrm{cm}^{-2}$).

\subsection{VII~Zw~292}

The \textit{Suzaku} results are reported here for the first time.
We obtain $\Gamma =
1.36_{-0.13}^{+0.14}
$ and $N_{\mathrm{H}}^{\mathrm{LOS}} =
41.14_{-2.49}^{+2.71}
\times 10^{22} \mathrm{cm}^{-2}$.
Using the \textit{XMM-Newton}/pn and \textit{Swift}/BAT data,
R17 obtained $\Gamma =
1.52 _{- 0.47 }^{+ 0.52 }
$ and $N_{\mathrm{H}}^{\mathrm{LOS}} =
61.7 _{- 19.0 }^{+ 38.3 }
\times 10^{22} \mathrm{cm}^{-2}$.
Similarly, \cite{Ursini18VIIZw292} obtained $\Gamma = 1.61\pm{0.17}$
and $N_{\mathrm{H}}^{\mathrm{LOS}} =40\pm{8} \times 10^{22}
\mathrm{cm}^{-2}$ from the \textit{XMM-Newton}/pn, \textit{Swift}/BAT,
and \textit{NuSTAR} data. These results are consistent with ours.

\subsection{3C~403}

We obtain $\Gamma =
1.87_{-0.24}^{+0.19}
$ and $N_{\mathrm{H}}^{\mathrm{LOS}} =
43.65_{-6.71}^{+7.1}
\times 10^{22} \mathrm{cm}^{-2}$.
These values agree with those of R17 ($\Gamma = 1.69_{-0.13}^{+0.14}
$ and $N_{\mathrm{H}}^{\mathrm{LOS}} = 49.0_{-4.3}^{+17.1}
\times 10^{22} \mathrm{cm}^{-2}$) based on the \textit{Swift}/XRT
and \textit{Swift}/BAT data within the errors.
Fitting the same \textit{NuSTAR} spectra as ours
with the \textsc{pexrav} model, 
\cite{Panagiotou20} obtained $\Gamma =1.71_{-0.28}^{+0.09}$,
and $N_{\mathrm{H}}^{\mathrm{LOS}}=34.6_{-5.3}^{+3.2} \times 10^{22}
\mathrm{cm}^{-2}$, 
which are consistent with our results. However, 
\cite{Tazaki11} obtained a somewhat flatter slope, $\Gamma =
1.53_{-0.02}^{+0.03}$, and $N_{\mathrm{H}}^{\mathrm{LOS}}
=61_{-5}^{+6} \times 10^{22} \mathrm{cm}^{-2}$,
by fitting the \textit{Suzaku} and \textit{Swift}/BAT spectra with an
analytic model including \textsc{pexrav}.
We infer that it may be due to possible spectral variability
between the \textit{NuSTAR} and \textit{Suzaku} observation epochs
as indicated by the very small value of the time-variability constant
(0.22, Table~3).

\subsection{3C~105}

We obtain $\Gamma =
1.48_{-0.12}^{+0.12}
$ and $N_{\mathrm{H}}^{\mathrm{LOS}} =
39.27_{-3.58}^{+2.97}
\times 10^{22} \mathrm{cm}^{-2}$.
These values are consistent with those of R17 based on
the \textit{XMM-Newton}/pn and \textit{Swift}/BAT spectra ($\Gamma =
1.29_{-0.29}^{+0.31}
$ and $N_{\mathrm{H}}^{\mathrm{LOS}} =
43.7_{-5.6}^{+7.6}
\times 10^{22} \mathrm{cm}^{-2}$).
\cite{Fioretti13} obtained $\Gamma =
1.78_{-0.19}^{+0.2}
$ and $N_{\mathrm{H}}^{\mathrm{LOS}} =
45.96_{-6.56}^{+6.24}
\times 10^{22} \mathrm{cm}^{-2}$ using \textit{Suzaku} and
\textit{Swift}/BAT data, which are also
consistent with ours.

\subsection{3C~452}

Our best-fit values, $\Gamma =
1.61_{-0.07}^{+0.06}
$ and $N_{\mathrm{H}}^{\mathrm{LOS}} =
43.39_{-5.78}^{+2.24}
\times 10^{22} \mathrm{cm}^{-2}$, are consistent
with
\citet{Fioretti13}, who obtained $\Gamma =1.55_{-0.11}^{+0.14}$
and $N_{\mathrm{H}}^{\mathrm{LOS}} =
43.52_{-6.92}^{+10.85}
\times 10^{22} \mathrm{cm}^{-2}$ using \textit{Suzaku} and
\textit{Swift}/BAT data.
\cite{Panagiotou20}
analyzed the same \textit{NuSTAR} spectra as ours and derived
$N_{\mathrm{H}}^{\mathrm{LOS}}=46.8\pm 2.3 \times 10^{22} \mathrm{cm}^{-2}$
with the \textsc{pexrav} model by fixing $\Gamma$ at 1.73.
This value is consistent with ours.
Using the \textit{XMM-Newton}/pn and \textit{Swift}/BAT data,
R17 reported a flatter slope and a larger absorption
($\Gamma=1.31_{-0.18}^{+0.2}$ and
$N_{\mathrm{H}}^{\mathrm{LOS}}= 57.5_{-9.7}^{+10.1}
\times10^{22}\mathrm{cm}^{-2}$) than our results.
This is most likely due to time variability between the
\textit{XMM-Newton} and \textit{Suzaku} observations, because we find
that soft X-ray flux below $\sim$3 keV was significantly lower in the former
epoch than in the latter one.

\subsection{PKS~2356--61}

Our results ($\Gamma =
1.58_{-0.01}^{+0.05}
$ and $N_{\mathrm{H}}^{\mathrm{LOS}} =
13.64_{-0.34}^{+0.56}
\times 10^{22} \mathrm{cm}^{-2}$)
are consistent with those of R17 obtained from the \textit{Suzaku} and
\textit{Swift}/BAT data ($\Gamma =
1.59_{-0.16}^{+0.13}
$ and $N_{\mathrm{H}}^{\mathrm{LOS}} =14.5_{-0.7}^{+0.7}
\times 10^{22} \mathrm{cm}^{-2}$).
Using the same \textit{NuSTAR} data as ours and
the \textit{Swift}/BAT data,
\cite{Ursini18PKS2356-61} obtained $\Gamma =
1.7\pm{0.3}$ and $N_{\mathrm{H}}^{\mathrm{LOS}} =
14\pm{5} \times 10^{22} \mathrm{cm}^{-2}$,
which are also consistent with our results.
\cite{Panagiotou20} also analyzed 
the same \textit{NuSTAR} data and 
obtained $\Gamma= 1.41\pm 0.27$ and
$N_{\mathrm{H}}^{\mathrm{LOS}}=12.5_{-4.2}^{+4.3} \times 10^{22}
\mathrm{cm}^{-2}$ by applying the \textsc{pexrav} model.
These values are in good agreement with ours.

\bsp    
\label{lastpage}
\end{document}